\documentclass[preprint,12pt]{elsarticle}

\usepackage{amssymb}
\usepackage[utf8]{inputenc}

\usepackage{algpseudocode}
\usepackage{graphicx}
\usepackage{amsmath}
\usepackage[version=4]{mhchem}
\usepackage{siunitx}
\usepackage{longtable,tabularx}
\usepackage{algorithm}
\usepackage{caption}
\usepackage{subcaption}
\usepackage{tikz}
\usepackage{array,multirow}
\usepackage[mathscr]{euscript}
\usepackage{pifont}

\journal{computer and fluids}

\begin{document}

\begin{frontmatter}



\title{On The Implicit Large Eddy Simulation of Turbomachinery Flows Using The Flux Reconstruction Method}


\author[inst1,inst2]{Feng Wang}

\affiliation[inst1]{organization={School of Aeronautics and Astronautics},
            addressline={Shanghai Jiao Tong University}, 
            country={People's Republic of China}}
\affiliation[inst2]{organization={Oxford Thermofluids Institute, Department of Engineering Science},
            addressline={University of Oxford}, 
            country={United Kingdom}}



\begin{abstract}
A high-order flux reconstruction solver has been developed and validated to perform implicit large-eddy simulations of industrially representative turbomachinery flows. The T106c low-pressure turbine and VKI LS89 high-pressure turbine cases are studied. The solver uses the Rusanov Riemann solver to compute the inviscid fluxes on the wall boundaries, and HLLC or Roe to evaluate inviscid fluxes for internal faces. The impact of Riemann solvers is demonstrated in terms of accuracy and non-linear stability for turbomachinery flows. It is found that HLLC is more robust than Roe, but both Riemann solvers produce very similar results if stable solutions can be obtained. For non-linear stabilization, a local modal filter, which combines a smooth indicator and a modal filter, is used to stabilize the solution. This approach requires a tuning parameter for the smoothness criterion.  Detailed analysis has been provided to guide the selection of a suitable value for different spatial orders of accuracy. This local-modal filter is also compared with the recent positivity-preserving entropy filter in terms of accuracy and stability for the LS89 turbine case. The entropy filter could stabilize the computation but is more dissipative than the local modal filter. Regarding the spanwise spacing of the grid, the case of the LS89 turbine shows that a $z^+$ of approximately $45 - 60$ is suitable for obtaining a satisfactory prediction of the heat transfer coefficient of the mean flow. This would allow for a coarse grid spacing in the spanwise direction and a cost-effective ILES aerothermal simulation for turbomachinery flows. 
\end{abstract}

\begin{graphicalabstract}
\includegraphics[height=9cm]{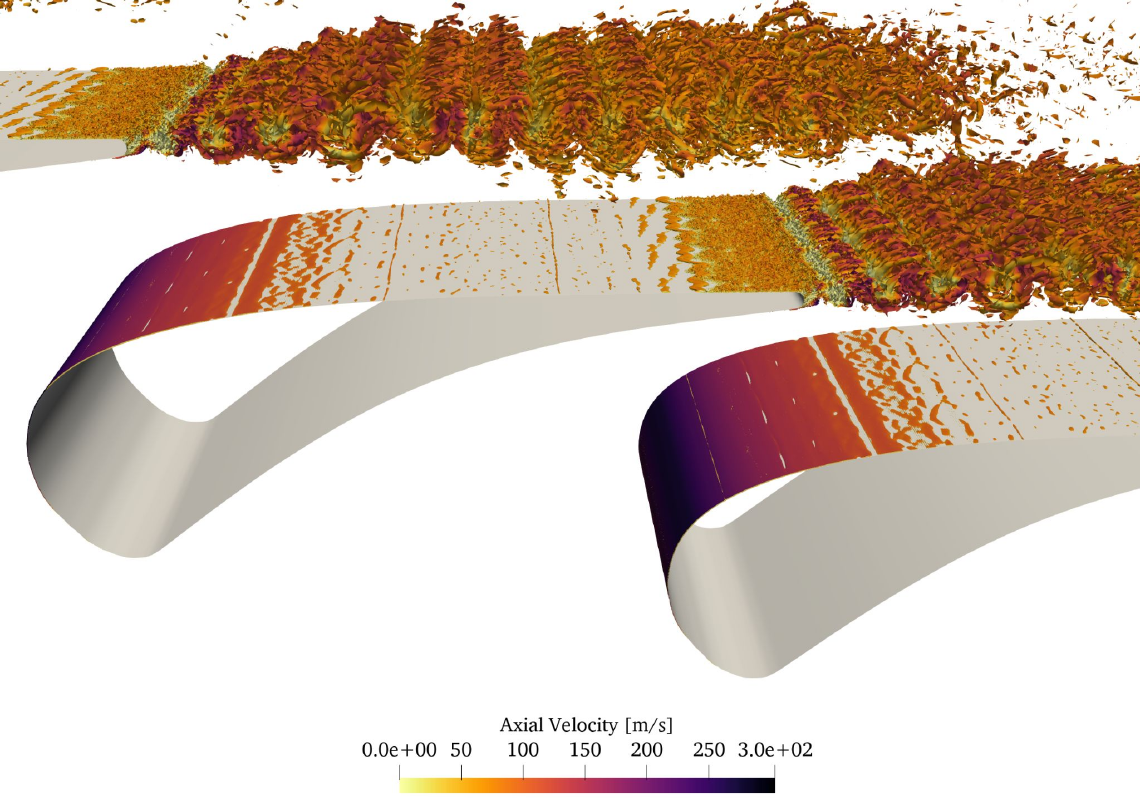}
\end{graphicalabstract}

\begin{highlights}
\item Effect of Riemann solvers on accuracy and numerical stability is investigated for turbomachinery flows.
\item A local modal filter is developed to stabilize the computation, and detailed analysis is provided to guide the selection of a suitable smoothness criterion.
\item The positivity preserving entropy filter is compared with the local modal filter and assessed in turbomachinery flows.
\item Relatively large spanwise grid spacing can be used to obtain mean flow quantities with satisfactory agreement with experimental data.
\end{highlights}


\end{frontmatter}


\section{Introduction}
Computational Fluid Dynamics (CFD) has progressed significantly in the last few decades and has transformed the design process in the aerospace industry~\cite{cfd_2030}. With the advancement of computing power and numerical algorithms, scale-resolving simulations, such as Large Eddy Simulation (LES), start complementing not only the experimental campaigns but also low-order numerical tools at the design stage. Reynolds-averaged Navier-Stokes (RANS) simulations and unsteady RANS (URANS) are the current workhorse in the industry for aerothermal design. However, it is well known that the RANS/URANS approach can face difficulties in reliably predicting vortex-dominated and separated flows and its associated heat transfer phenomenon.

The use of discontinuous spectral element methods (DSEM) has grown in prevalence in the past few years for LES due to their ability to achieve high-order spatial accuracy in unstructured grids and a compact stencil that is suited for massively parallel computations. Popular DSEM methods are the nodal Discontinuous Galerkin (DG) approach~\cite{dg_reed,dg_cockburn_shu} and, more recently, the Flux Reconstruction (FR)~\cite{Huynh2007,Huynh2009} or the Correction Procedure via Reconstruction (CPR)~\cite{Wang2009} method. Despite the differences in these two methods, the numerical procedures developed in the collocation-based nodal DG could be applied to FR/CPR (such as filtering and boundary treatment) with little modification and vice versa. Implicit LES (ILES) is frequently used in DSEM, as the built-in dissipation of DSEM is found to be adequate as a subgrid model.  The superior performance of DSEM over traditional second-order finite-volume solvers on ILES has been demonstrated by Vermeire et al.~\cite{Vermeire2017} and Jia and Wang~\cite{Jia2019}.

DSEM requires a Riemann solver to calculate the inviscid flux on the element interface, such as the Rusanov~\cite{Rusanov1962}, Roe~\cite{Roe1981} or HLLC~\cite{Toro2009-bg} Riemann solvers. Because these approximate Riemann solvers contain numerical dissipation, this can have an impact not only on the accuracy but also on the numerical stability of the computation. Beck et al.~\cite{Beck2013} and Moura et al.~\cite{moura_ic_report} demonstrated such an effect on simple flows (e.g. Taylor Green vortex), but for industry-representative turbomachinery flows, such an effect has not been studied in a systematic manner. Furthermore, for wall-bounded internal flows, it was found that the treatment of convective fluxes on wall boundaries can have a non-negligible impact on the robustness of the solver, especially for under-resolved cases~\cite{Mengaldo2014}.

Industrial turbomachinery flows typically have large Reynolds numbers, and this leads to a wide spectrum of time and length scales in turbulent flows. It would be prohibitively expensive to resolve all these scales. Therefore, with respect to the true flow physics,  it is inevitable that turbulent flows will be under-resolved. Collocation projection is commonly used in DSEM due to its computational efficiency. However, due to the non-linearity of the Navier-Stokes equations, this collocated approach can generate errors if any of the modes arising from the nonlinear terms are outside the span of the set of basis functions. In these cases, the energy from the unresolved (i.e. under-integrated) modes is aliased onto the lower modes and leads to aliasing errors. This could consequently cause numerical instability~\cite{Jameson2011,Kirby2003,Gassner2012}. Over-integration~\cite{Spiegel2015}, which is also referred to as polynomial de-aliasing, could be used to alleviate this problem. This is accomplished by using a set of quadrature points to increase the sampling of the flux function above what is capable by the solution points. But there is evidence~\cite{Spiegel2015} that this may not eliminate all instabilities in an underresolved case. Various techniques have been developed to stabilize DSEM simulations, such as artificial viscosity~\cite{VonNeumann1950,Karamanos2000,Persson2006,Kirby2006,Barter2010}, limiting~\cite{Zhang2010,Zhong2013,Li2017,zjWang2017}, using a split form~\cite{Gassner2016}, filtering~\cite{Hesthaven2007-yn}, etc. 

Filtering is an attractive stabilization technique due to its ease of implementation and computational efficiency. However, care must be taken when filtering is applied to DSEM, as it can significantly undermine the accuracy of DSEM~\cite{Park2017}. The amount of filtering should be applied as much as needed to stabilize the solution but also as little as possible to avoid excessive dissipation. A modal filter could be localized using a smooth indicator so that the smooth-flow region is not subject to filtering. However, this strategy requires a smoothness criterion to determine whether the element should be filtered or not. This method has not been systematically studied in the literature for turbomachinery flows, and the choice of the smooth criterion is also not clear. The first contribution of this paper is: we will demonstrate the performance of this local modal filtering approach on turbomachinery flows and perform a detailed analysis of the smoothness criterion to provide a guide to select a suitable value. The recent entropy filter of Dzanic and Witherden~\cite{dzanic_filter} can also be considered as a localized modal filtering approach. Physical constraints are used as a ``smoothness" criterion to determine which element should be filtered. This entropy filter is implemented and its performance is assessed in turbomachinery flows and compared with the local modal filter. 

Riemann solvers have been found to have an impact on accuracy and numerical stability on simple turbulent flows~\cite{moura_ic_report} using DSEM. These effects have not been systematically studied for turbomachinery flows. In this work, such a study will be performed on industrially representative turbomachinery cases. In addition, Alhawwary and Wang~\cite{Alhawwary2019} performed a study of the spanwise grid spacing on the prediction of mean flow variables (i.e., isentropic Mach number) in the T106c turbine using CPR. However, there have been no similar studies on the prediction of heat transfer coefficients in turbine blades, which is more challenging to predict than the pressure distribution on the blade surface. In this paper, we demonstrate this effect using the VKI LS89 turbine case. These are the second contribution of this work.

The computations in this paper are performed with a new in-house FR solver, AeroThermal High-Order Simulation (ATHOS). This paper is organized as follows. The governing equations of compressible flow and the FR method are briefly described. Then the local modal filter is introduced. The performance of the local modal filter on Taylor Green vortex, T106c, and LS89 turbine cases is demonstrated, the effect of Riemann solvers on accuracy and numerical stability is analyzed, and a detailed analysis of the smoothness criterion of the modal filter is presented.  This is then followed by the conclusions and future work.  

\section{Methodology}
\subsection{Governing  Equation}
The Navier-Stokes equations for compressible flows in the differential form can be written as:
\begin{equation}
\label{eqn::ns_equation}
\frac{\partial \mathbf{U}}{\partial t} + \mathbf{\nabla} \cdot \mathbf{F}(\mathbf{U}, \mathbf{\nabla}\mathbf{U}) = 0
\end{equation}
in which $\mathbf{U}$ is the conservative variable $\mathbf{U} = ({\rho, \rho u, \rho v, \rho w, \rho e})$, $\rho$ is the density of the fluid, $\mathbf{V}=(u,v,w)$ is the velocity vector in the Cartesian coordinate system, $\mathbf{e}$ is the total energy per unit mass. For a perfect gas, the pressure can be related to the total energy $e$ as: 

\begin{equation}
  p = \rho (\gamma - 1) (e - \frac{1}{2}  ||\mathbf{V}||^2)
\end{equation}
Where $\gamma$ is the specific heat ratio and a value of 1.4 is used by default in this study. 

The flow variable $\mathbf{U}$ and its gradient $\mathbf{\nabla} \mathbf{U}$ are required to evaluate the flux $\mathbf{F}$ and the flux can be written as $\mathbf{F} = \mathbf{F}^I - \mathbf{F}^v$. The inviscid flux $\mathbf{F}^I$ reads:

\begin{align}
\mathbf{F}^I = 
  \begin{pmatrix}
  \rho u & \rho v & \rho w\\
  \rho u^2+p & \rho uv & \rho u w \\
  \rho uv & \rho v^2 + p & \rho v w \\
  \rho uw & \rho vw & \rho w^2 + p \\
  \rho uh & \rho vh & \rho wh \\
  \end{pmatrix}    
\end{align}
in which $h=e+\frac{p}{\rho}$ is the total enthalpy. The viscous flux $\mathbf{F}^v$ can be written as:
\begin{equation}
\label{eqn::ns_visc}
\mathbf{F}^v = 
  \begin{pmatrix}
  0 & 0 & 0\\
  \sigma_{xx} & \sigma_{yx} & \sigma_{zx} \\
  \sigma_{xy} & \sigma_{yy} & \sigma_{zy} \\
  \sigma_{xz} & \sigma_{yz} & \sigma_{zz} \\
  V_i\sigma_{ix} - q_x & V_i\sigma_{iy} - q_y & V_i\sigma_{iz} - q_z \\
  \end{pmatrix}    
\end{equation}
In the viscous flux formulation, $\sigma_{ij}$ is the viscous stress and for a Newtonian fluid it can be written as:
\begin{equation}
    \sigma_{ij} = \mu (\frac{\partial u_i}{\partial x_j} + \frac{\partial u_j}{\partial x_i}) - \frac{2}{3} \mu \delta_{ij} \frac{\partial u_k}{\partial x_k}
\end{equation}
in which $\mu$ is the dynamic viscosity of the fluid. $\delta_{ij}$ is  the Kronecker delta. $q_i$ is the heat flux and it reads:
\begin{equation}
    q_i = - k \frac{\partial T}{\partial x_i}
\end{equation}
The coefficient of thermal conductivity $k$ is can be computed as:
\begin{equation}
    k = \frac{C_p} \mu{Pr}
\end{equation}
in which $Pr$ is the Prandtl number, Cp is the specific heat at constant pressure and it can be computed as $Cp = \frac{\gamma}{\gamma -1} R T$, $R$ is the gas constant. In this work, the following values are used for these constants for the ideal gas: $Pr = 0.72$, $\gamma = 1.4$ and $R = 287 [\text{J}/(kg \cdot \text{K})]$. Finally, the dynamic viscosity $\mu$ is computed using the Sutherland's law as:

\begin{equation}
    \mu = \mu_{\text{ref}} (\frac{T}{T_{\text{ref}}})^{1.5} \frac{T_{\text{ref}} + S}{T + S}
\end{equation}
in which $T_{\text{ref}}$ is $291.5 [\text{K}]$, $S$ is $120 [\text{K}]$, $\mu_{\text{ref}}$ is $1.827 \times 10^{-5} [kg/(ms)]$

\subsection{High-Order Flux Reconstruction}
The FR method is an efficient discontinuous spectral element method to solve the partial differential equation (PDE) in differential form. It is briefly described here, and for a detailed description one can refer to Huynh~\cite{Huynh2007}, Vicent et al.~\cite{Vincent_2010} and Castonguay~\cite{Castonguay_thesis}.  The NS equation of the compressible flow (Equation~\ref{eqn::ns_equation}) is a second-order PDE. Before applying FR, it is first cast into a set of first order PDEs and this is shown in Equation~\ref{eqn::ns_eqn_1storder}~\cite{Hesthaven2007-yn}:
\begin{align}
    \label{eqn::ns_eqn_1storder}
    \frac{\partial \mathbf{U}}{\partial t} + \mathbf{\nabla} \cdot \mathbf{F}(\mathbf{U}, \mathbf{W}) &= 0 \nonumber \\
    \mathbf{W} - \mathbf{\nabla} \mathbf{\hat{U}} &= 0
\end{align}
where $\mathbf{W}$ is an auxiliary variable. The physical domain $\Omega$ is then decomposed into $N$ conformal and nonoverlapping subdomains in the first place. For convenience, Equation~\ref{eqn::ns_eqn_1storder} is mapped from the physical domain $\mathbf{x}=(x,y,z)$ domain to the computational space $\mathbf{\hat{x}}=(\xi, \eta, \zeta)$ as:
\begin{align}
    \label{eqn::ns_eqn_1storder_mapped}
    \frac{\partial \mathbf{\hat{U}}}{\partial t} + \mathbf{\hat{\nabla}} \cdot \mathbf{\hat{F}}(\mathbf{\hat{U}}, \hat{W}) &= 0 \nonumber \\
    \mathbf{\hat{W}} - \mathbf{\hat{\nabla}} \mathbf{U} &= 0
\end{align}
Where the hat symbol $\widehat{\cdot}$ represents the variables and derivatives in the computational space. The transformation between the physical space and the computational space is determined by the shape functions of the element and the resulting Jacobian matrix $J=\frac{\partial (x,y,z)}{\partial (\varepsilon, \eta, \zeta)}$. The details can be found in Zienkiewicz et al.~\cite{zienk_book} and Haga et al.~\cite{Haga2011}. 

\begin{figure}
\centering 
\includegraphics[height=8cm]{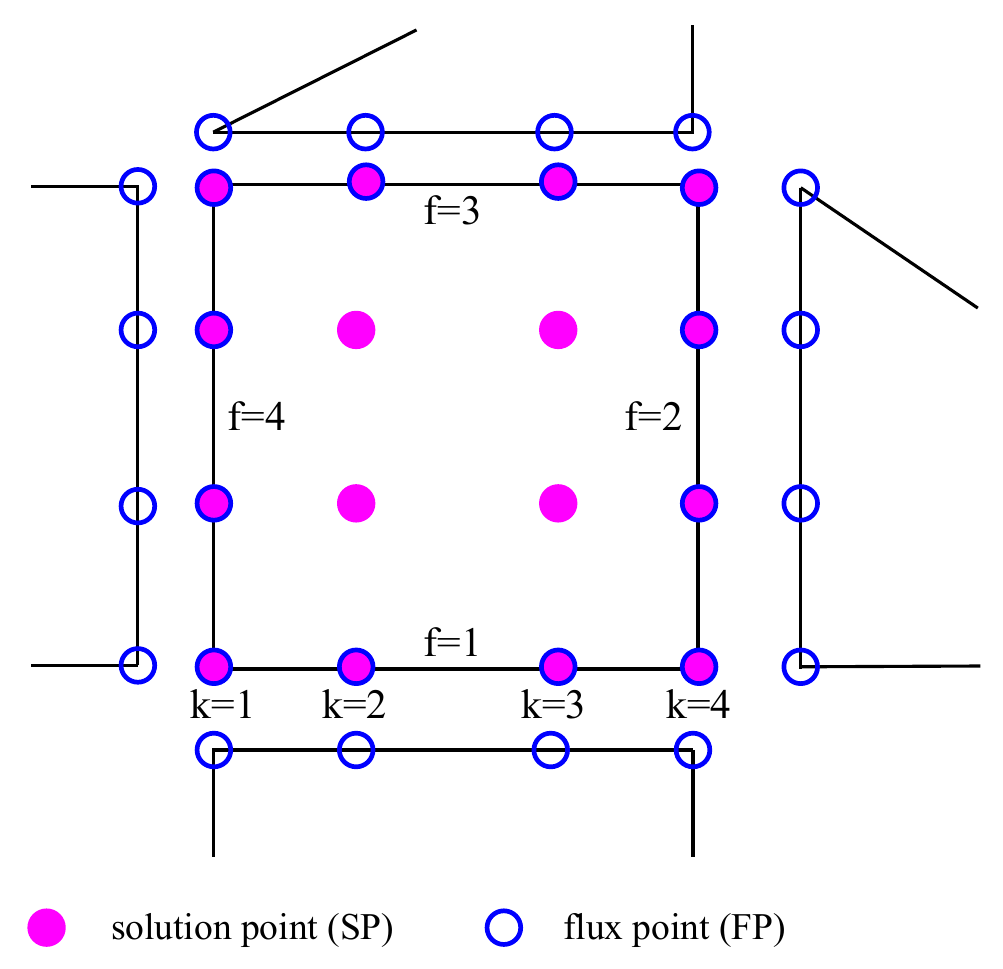}
\caption{Illustration of solution and flux points in FR.}
\label{fig:sp_fp}       
\end{figure}

In the computational domain, the flow variable $\mathbf{\hat{U}}$ is represented by a multidimensional polynomial of degree $p$. The polynomial is defined by a set of solution points in the element, and the representation of $\mathbf{\hat{U}}$ takes the form:

\begin{equation}
   \mathbf{\hat{U}} = \sum^{N_{sp}}_{i=1} \mathbf{\hat{U}_i} L_i(\mathbf{\hat{x}})
\end{equation}
where $\mathbf{\hat{U}_i}$ is the value of the transformed conservative variable on the $i^{\text{th}}$ solution point, $L_i(\mathbf{\hat{x}})$ is the multidimensional Lagrange polynomial associated to the $i^{\text{th}}$ solution point, $N_{sp}$ is the number of solution points in this element. In this work, the Legendre-Gauss-Labbato (LGL) point is used for the solution point and this is illustrated in Fig.~\ref{fig:sp_fp}.

For the flux $\mathbf{\hat{F}}$ in Equation~\ref{eqn::ns_eqn_1storder}, it can be written as the sum of a discontinuous component $\mathbf{\hat{F}}^D$ and a correction term $\mathbf{\hat{F}}^C$ as follows:

\begin{equation}
    \mathbf{\hat{F}} = \mathbf{\hat{F}}^D + \mathbf{\hat{F}}^C
\end{equation}

The discontinuous component $\mathbf{\hat{F}}^D$ is computed using a collocated projection approach, and it can be represented by a polynomial of degree $p$ as: 

\begin{equation}
   \label{eqn::flux_collocate}
   \mathbf{\hat{F}} = \sum^{N_{sp}}_{i=1} \mathbf{\hat{F}_i} L_i(\mathbf{\hat{x}})
\end{equation}

in which $\mathbf{\hat{F}_i}$ is the transformed flux computed on the $i^{\text{th}}$ solution point. They are computed using the flow variables and their gradients stored on the $i^{\text{th}}$ solution point. $L_i(\mathbf{\hat{x}})$ is the same multidimensional Lagrange polynomial that is used for the flow variable. To calculate the correction component $\mathbf{\hat{F}}^C$, a set of flux points is defined on the boundary of the element, and they are used to couple the adjoining element. This is illustrated by the blue point in Fig.~\ref{fig:sp_fp}. The correction component $\mathbf{\hat{F}}^C$ can be constructed as:

\begin{equation}
  \mathbf{\hat{F}}^C(\mathbf{\hat{x}}) = \sum^{N_f}_{f=1} \sum^{N^f_{fp}}_{k=1} [ (\mathbf{\hat{F}} \cdot \mathbf{\hat{n}})^I_{f,k} - (\mathbf{\hat{F}} \cdot \mathbf{\hat{n}})^D_{f,k} ] \mathbf{g}_{f,k}(\mathbf{\hat{x}})
\end{equation}
in which $f$ is the index of the element boundary, $N_f$ is number of boundary interfaces of this element, $k$ is the index of the flux point on the $f^{\text{th}}$ element boundary, and $N^f_{fp}$ is the number of flux points on the interface $f$, $(\mathbf{\hat{F}} \cdot \mathbf{\hat{n}})^D_{f,k}$ is the discontinuous flux on the flux point $(f,k)$.  $(\mathbf{\hat{F}} \cdot \mathbf{\hat{n}})^I_{f,k}$ is a normal transformed numerical flux calculated at the flux point $(f,k)$. 
$\mathbf{g}_{f,k}$ is the correction function associated with the flux point $(f,k)$ and its specific form determines various properties of the resulting FR scheme~\cite{Vincent_2010,Vincent_2015}. It satisfies the following property:

\begin{equation}
  \label{eqn::correct_func}
  \mathbf{g}_{f,k} \cdot \mathbf{\hat{n}}_{f2,k2}
   = \begin{cases}
    1, & \text{if} f=f_2 , k=k_2 \\
   0, & \text{otherwise}.
  \end{cases}
\end{equation}
in which $f_2$ and $k_2$ are indices of the element boundary and the flux point on a specific element boundary. In this work, the correction function is chosen in such a way that the nodal discontinuous Galerkin approach is recovered~\cite{Huynh2007}.

With respect to the common numerical flux $(\mathbf{\hat{F}} \cdot \mathbf{\hat{n}})^I_{f,k}$ on the element interface, for the invsicid contribution,  the flow solutions on both side of the interface, namely $\mathbf{U}^+_{f,k}$ and $\mathbf{U}^-_{f,k}$, are used to solve a Riemann problem using an approximate Riemann solver, such as Rusanov~\cite{Rusanov1962}, Roe~\cite{Roe1981} and HLLC~\cite{Toro2009-bg}. Since the Riemann solver inherently contains numerical dissipation, for example, the Rusanov solver can be considered as a central-difference flux using a scalar dissipation (i.e. using the largest local wave speed as the scalar), while the Roe solver can be considered as a central-difference flux using a matrix dissipation. Therefore, the choice of Riemann solver can have an impact on the accuracy and robustness of implicit LES with DSEM~\cite{moura_ic_report}. 

To calculate the viscous contribution, the local discontinuous Galerkin (LDG)~\cite{Cockburn1998,Castonguay2013} approach is used. This can be written as:

\begin{equation}
    \mathbf{F}^{v,\text{LDG}}_{f,k} = \frac{\mathbf{F}^v(\mathbf{U}^+_{f,k}) + \mathbf{F}^v(\mathbf{U}^-_{f,k})}{2} + \tau(\mathbf{U}^-_{f,k} - \mathbf{U}^+_{f,k}) + \beta (\mathbf{F}^v(\mathbf{U}^-_{f,k}) - \mathbf{F}^v(\mathbf{U}^+_{f,k}))
\end{equation}
in which $\mathbf{F}^v(\mathbf{U}^{\pm}_{f,k})$ is the viscous flux (see Equation~\ref{eqn::ns_visc}) using flow variable  $\mathbf{U}^{\pm}_{f,k}$, $\tau$ is a penalty parameter controlling the jump in the solution, in this work, $\tau = 0.0$, and this corresponds to the minimum dissipation version of LDG. $\beta$ is the directional parameter and has the value of $\pm0.5$, whose sign is decided locally to ensure a biased either upwind or downwind direction~\cite{Castonguay2013,Hesthaven2007-yn}.


FR solves the differential form of the NS equation. The divergence of the discontinuous flux $\mathbf{\hat{F}}^D$ and the correction flux  $\mathbf{\hat{F}}^C$ are required. For the discontinuous flux, the divergence of the Lagrange polynomial $L_i(\mathbf{\hat{x}})$ needs to be computed, while for the correction component, the divergence of $\mathbf{g}_{f,k}$ is evaluated~\cite{Castonguay_thesis}. Finally, in terms of time integration,  an explicit low-storage fourth-order ﬁve-stage Runge–Kutta method is used to advance the solution in time~\cite{rk45}.

\subsection{Stabilization with filtering}
DSEM methods can face robustness issues, and stabilization techniques are usually required to enhance robustness to handle underresolved flows or strong discontinuity (e.g., shocks). In this work, a local modal filtering (LMF) approach is developed. The general concept is to apply the modal filter in regions where stabilization is required while leaving the smooth-flow region untouched. Modal filtering is a simple and classical approach in DSEM and it can be represented as a matrix operation for the nodal solutions as~\cite{Hesthaven2007-yn}:

\begin{equation}
    \mathscr{F} = \mathbf{\mathscr{V}} \Lambda \mathscr{V}^{-1}
\end{equation}
in which $\mathscr{F}$ is the filtering matrix,  $\mathbf{\mathscr{V}}$ is the Vandermonde matrix and $\mathbf{\Lambda}$ is the diagonal matrix and it has entries:

\begin{equation}
  \label{eqn::modal_filter}
\Lambda_{ii} = \sigma(\eta_i)   = 
   \begin{cases}
    1, & \eta_i < \eta_c \\
   e^{-\alpha (\frac{\eta_i - \eta_c}{1-\eta_c})^s}, & \eta_c < \eta_i < 1
  \end{cases}
\end{equation}
where $\eta_i = \frac{i}{N_p - 1}$, $i$ is the mode index and $N_p$ is the maximum order of the polynomial. $\eta_c$ is the cutoff mode threshold, below which the modes will not be filtered. $\alpha$ and $s$ determine the strength of the filter. $s$ is an even number and its values usually range from 2 to 8. $s$ is related to the scale selectivity of the filter and in practice a lower value of $s$ normally leads to a stronger filter. The value of $\alpha$ is $-\log_{10}(\varepsilon_m)$ and $\varepsilon_m$ is the machine zero~\cite{Hesthaven2007-yn}. For simulations with double precision $\alpha \approx 36$. It is obvious that the power of the exponent in Equation~\ref{eqn::modal_filter} is a negative number; therefore, the modal filter is always dissipative.

Applying the modal filter~\ref{eqn::modal_filter} directly to the whole flow field can significantly undermine the accuracy of a DSEM solution~\cite{Park2017}. Therefore, it should only be applied when it is needed. Therefore, the concept of a local modal filter is developed here. To achieve this aim, a smooth indicator is required to shield the smooth flow region from modal filtering. In this work, the smooth indicator proposed by Persson and Peraire~\cite{Persson2006} is used. Within each element $\Omega_e$, a smooth indicator $s_e$ is used to assess the smoothness of the flow and is defined as:

\begin{equation}
    \label{eqn::lmf_smooth_indicator}
    s_e = \log_{10} \frac{(\phi - \hat{\phi})_e}{(\phi, \phi)_e}
\end{equation}
where $(\cdot, \cdot)_e$ is the standard inner product in $L_2(\Omega_e)$. $\phi$ is a flow variable that is used to assess the smoothness of a flow. With respect to the choice of flow variable $\phi$, according to Persson and Peraire~\cite{Persson2006}, density $\rho$ is suggested. It is possible that other flow variables could be used, such as total internal energy, such a study will be reported in future publications. $\hat{\phi}$ is a truncated polynomial expansion of $\phi$, but only contains expansions up to order $p-1$. $s_e$ is then compared with certain criteria to determine the smoothness of the flow. 

Li and Wang~\cite{Li2017} used the smooth indicator of Persson and Peraire for their limiter-based stabilization technique to prevent the limiter from being used in the smooth flow region. In their work, the smoothness of the flow can be categorized as shown in Table~\ref{tab::smooth_indicator}. In the table, $\kappa$ is a tuning parameter and represents the width of the activation "ramp", a value of 4 is used in the work of Li and Wang. $s_0=-C_0 \ln(p)$, where $p$ is the degree of polynomial. The value of $C_0 = 3$ is suggested by Li and Wang, but according to Klockner et al.~\cite{Klckner2011}, $C_0 = 4$. In this work, the value of the latter is used. For different polynomial orders (i.e., 2 to 3), the values of $s_0 - \kappa$, $s_0$, and $s_0 + \kappa$ are tabulated in Table~\ref{tab::smooth_indicator_value} and can be potential criteria for determining flow smoothness.

\begin{table}
\centering
\caption{Category of flow smoothness}
\label{tab::smooth_indicator}
\begin{tabular}{ c c }
\hline \hline
 $s_e < s_0 - \kappa$ & smooth  \\ \hline 
 $s_0 - \kappa \leq s_e \leq s_0 + \kappa$ & intermediate  \\  \hline
 $s_e > s_0 + \kappa$ & not smooth \\ \hline \hline
\end{tabular}
\end{table}

\begin{table}
\centering
    \caption{Comparative study of flow smoothness criteria $s_0$}

	\begin{subtable}[h]{0.45\textwidth}
    \caption{Tabulated value of $s_0$ flow smoothness criteria from Li and Wang~\cite{Li2017}}
    \label{tab::smooth_indicator_value}
       \begin{tabular}{ c c c c}
        \hline \hline
         order & $ s_0 - \kappa$ & $s_0$ & $s_0 + \kappa$  \\ \hline 
         p2   &    -6.7             &    -2.7   &   2.77       \\ \hline
         p3   &    -8.3             &    -4.3   &  -0.3       \\ \hline \hline
        \end{tabular}
	\end{subtable}
    \hfill
 	\begin{subtable}[h]{0.45\textwidth}
      \caption{Potential values for $s_0$ to apply the modal filter locally}
      \label{tab::smooth_indicator_value_new}
      \begin{tabular}{ c c c }
      \hline \hline
       order & $ s_0 - \kappa$ & $s_0$   \\ \hline 
       p2   &    -3.7             &    -2.7      \\ \hline
       p3   &    -5.3             &    -4.3       \\ \hline \hline
      \end{tabular}
	\end{subtable}
\end{table}

From the numerical experiment of the author, it is found that the values in Tabel~\ref{tab::smooth_indicator_value} cannot be used directly to localize the modal filter. Using $s_0 - \kappa$ leads to excessive dissipation, as considerably more elements are subject to filtering. The use of $s_0 + \kappa$ is found to not stabilize the solution. Using a value around $s_0$ is found to strike a balance between accuracy and robustness. As will be demonstrated in the following test cases, the range of $[s_0 - \kappa : s_0]$ offers good performance and $\kappa \approx 1$. This is summarized in Table~\ref{tab::smooth_indicator_value_new}. This range is used as a guide to select $s_0$ for the test cases. Posteriori analysis will also be performed to confirm the viability of this range.


The entropy filter (EF) proposed by Dzanic and Witherden~\cite{dzanic_filter} can be considered as a local modal filtering approach. For the modal filter in Equation~\ref{eqn::modal_filter}, if $s=2$ and $\eta_c=0$, the following second-order modal filter can be obtained:

\begin{equation}
  \label{eqn::modal_filter_ef}
  \Lambda_{ii} = \sigma(\eta_i) = e^{-\alpha \eta_i^2}
\end{equation}
in which $\eta_i = \frac{i}{N_p - 1}$, $i$ is the mode index and $N_p$ is the maximum order of the polynomial. The difference between EF and LMF is that the value of $\alpha$ in EF is determined automatically by examining certain physical principles, e.g. the discrete minimum entropy condition:

\begin{equation}
    \label{eqn::ef_s}
    s(\mathbf{U}(\mathbf{x}, t+ \Delta t)) \geq \min_{\Omega}(s(\mathbf{U}(\mathbf{x}, t))) - \varepsilon_s
\end{equation}
and the positivity-preserving principle for density and pressure:

\begin{equation}
    \rho \geq \rho_{\text{min}}, \quad p \geq p_{\text{min}}
\end{equation}
in which $s = \rho \ln(p \rho^{-\gamma})$ is the entropy, $\varepsilon_s$ is a tuning parameter and a value of $10^{-4}$ is used~\cite{Dzanic2022}. The minimum principle states that for each element, the entropy is non-decreasing in time in its domain of influence. For an explicit time-integration scheme, the CFL number is generally less than unity, and this domain constitutes the immediate neighbors of this element. The positivity-preserving principle makes sure that density and pressure are non-negative values and are forced to be larger than a minimum threshold, e.g. $\rho_{\min}=10^{-8}$ and $p_{\min} = 10^{-8}$. 

As presented in the previous text, a larger $\alpha$ will introduce more dissipation to the flow field. EF employs an optimization process to search for a suitable $\alpha$ that satisfies both the minimum entropy condition and the positivity-preserving principle for density and pressure. However, EF is more complicated to implement and also computationally more expensive than LMF, because it needs to search for a suitable $\alpha$ while LMF uses a user-specified value. The performance of EF will be assessed in the LS89 turbine case, and its performance will be compared with that of LMF.

Finally, it is noted that either LMF or EF is a solution-filtering stabilization technique and will introduce extra numerical dissipation \emph{locally} to the computation~\cite{Edoh2018}. Such numerical dissipation will contribute to the total numerical dissipation in the computation. This could have an impact on ILES as it relies on the numerical dissipation to act like a subgrid model. This impact will be studied in the following test cases, especially for the LS89 turbine case, which has a reasonably high Reynolds number and more elements are required to be filtered to stabilize the computation. On the other hand, Lamballais et al.~\cite{Lamballais_2021} showed that the solution filtering approach can be connected to the Spectral Vanishing Viscosity (SVV) technique in the finite difference framework, but such connection within the FR framework is not yet reported in the literature and is out of the scope of the current paper, this will be explored in the future work.        



\subsection{Boundary Condition}
\subsubsection{Inlet and Exit Boundaries}
The implementation of boundary conditions for the FR method is generally similar to a finite-volume code. For example, at the inlet, total pressure, total temperature, and flow angles are specified. At the exit, static pressure is prescribed.

\subsubsection{Wall Boundary}
Particular attention is required for the non-slip wall boundaries, where the flow gradient is high. Mengaldo et al.~\cite{Mengaldo2014} concluded that the approach that uses the values of ghost cells and a Riemann solver to compute the convective fluxes on the wall is the most robust.  This procedure is followed in this work. In our numerical experiment, the choice of a Riemann solver to compute the convective fluxes on the wall has a non-negligible impact on the solution stability near the wall. Three Riemann solvers are tested, namely Rusanov, Roe and HLLC. The most dissipative one, namely the Rusanov Riemann solver, is found to be the most robust one. This is consistent with the findings of Mengaldo et al.~\cite{Mengaldo2014}, although in their work the HLL and HLLC were compared, the HLL solver is found to be more robust than HLLC. 

However, it is preferable that other Riemann solvers (e.g. Roe and HLLC) could be used for the internal faces. Therefore, in the current implementation, a hybrid approach is used: the Rusanov Riemann solver is used to calculate the convective fluxes on the wall while another Riemann solver (i.e. Roe or HLLC) is used for the internal element interfaces. This hybrid approach is found to be robust and its performance will be demonstrated in the following test cases.

\section{Computational Framework}
The computations are performed with the newly developed in-house FR solver ATHOS. It is written in C++ and implements the FR method that was originally proposed by Huynh~\cite{Huynh2007} and was then further developed by Vincent et al.~\cite{Vincent_2010} and Castonguay~\cite{Castonguay_thesis}. Gauss-Lobatto-Legendre points are used for the solution and flux points (see Fig.~\ref{fig:sp_fp}). The solver currently supports linear tensor elements (such as linear quadrilateral and hexahedral elements), and a spatial accuracy of up to $p3$. The correction function is selected so that the collocation-based nodal DG scheme is recovered~\cite{Huynh2007}.  The overall description of the FR implementation is described in~\ref{sec:appendix_1}. 

\section{Results}
Three test cases are used and ILES is performed for all cases. The first case is the Taylor-Green vortex. The second case is the T106c low-pressure turbine. The third case is the VKI LS89 turbine. The first case is used to demonstrate that LMF does not introduce noticeable dissipation when the flow is smooth. The second case demonstrates the performance of LMF in a moderate Reynolds number (i.e.,Re=80000) case and the effect of Riemann solvers. The third case is more challenging. It demonstrates the performance of LMF on stabilizing a turbine case with heat transfer at an industrially representative Reynolds number (i.e., Re=$1.13\times 10^6$). The performance of EF is assessed, and the impact of Riemann solver and spanwise grid spacing is also discussed.

\subsection{Taylor Green Vortex at Re=1600 and Re=5000}
The subsonic Tayor-Green vortex at Reynolds numbers 1600 and 5000 and Mach number 0.1 is used to verify that LMF does not introduce extra dissipation into the computation in the absence of strong discontinuity (i.e. shocks). The problem is solved in a periodic domain with size $[0:2 \pi]$ in each direction. The velocity and pressure in this periodic domain are initialized as:
\begin{equation}
  \begin{bmatrix}
  u(\mathbf{x},0) \\
  v(\mathbf{x},0) \\
  w(\mathbf{x},0) \\
  p(\mathbf{x},0) \\
  \end{bmatrix} 
  =
  \begin{bmatrix}
  \sin(x) \cos(y) \cos(z) \\
  -\cos(x) \sin(y) \cos(z) \\
  0 \\
  p_0 + \frac{1}{16} (\cos(2x) + \cos(2y))\cos(2z+2) \\
  \end{bmatrix}    
\end{equation}
where $p_0 = 100 \text{Pa}$. In the following text, the SI unit is used by default. The dynamic viscosity $\mu$ is set to the value of $1.827 \times 10^{-5} [kg/(ms)]$ and density $\rho$ is adjusted to obtain the Reynolds number of 1600 and 5000, respectively.  The Reynolds number is computed as $\text{Re} = \frac{\rho V_0 L}{\mu}$. $V_0$ is $0.1 \sqrt{p/\rho}$ and $L=1 m$. 

The kinetic energy of the flow is integrated in the domain and it is denoted as $K$. The dissipation rate of $K$ with time, $\varepsilon_{K}=\frac{\mathrm{d} K}{\mathbf{d} t}$ is then monitored to assess the accuracy of the numerical scheme.  The domain is meshed with a structured mesh with the size of $64\times64\times64$, and the grid spacing is uniform in all directions. A $p3$ solution is computed and the HLLC Riemann solver is used. With respect to the effect of Riemann solver on this case readers can refer to Moura et al.~\cite{moura_ic_report}. The LMF is used and $s_0$ is set to the value of $-4$.  

\begin{figure}
\centering 
\includegraphics[height=6cm]{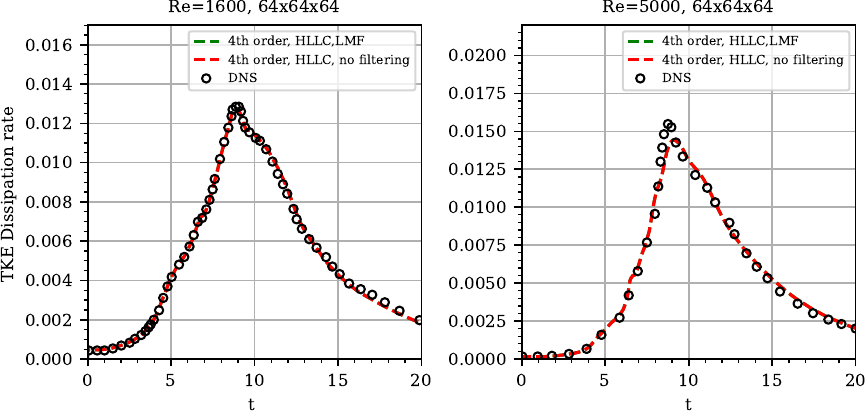}
\caption{Dissipation rate of turbulent kinetic energy with LMF at Re=1600 and 5000.}
\label{fig:tgv}       
\end{figure}

Figure~\ref{fig:tgv} shows the predicted time evolution of $\varepsilon_{K}$ using LMF at Reynolds numbers of 1600 and 5000, respectively. The results are compared with DNS data. Time $t$ is the normalized time, which is computed as $t=t_{\text{physical}}/t_c$. $t_c$ is the characteristic time and it is defined as $L/V_0$.  At Re=1600, the DNS data is from Gassner and Beck~\cite{Gassner2012}. For Re=5000 it is computed by Dairay et al.~\cite{Dairay2017} using the Incompact3d code. From the comparison, it can be seen that there is no noticeable difference between the results of LMF and the one without filtering. Both results show excellent agreement with the reference DNS data. This confirms that LMF does not introduce noticeable dissipation to the simulation in the absence of strong discontinuity in the flow.

\begin{figure}
\centering 
\includegraphics[height=4.5cm]{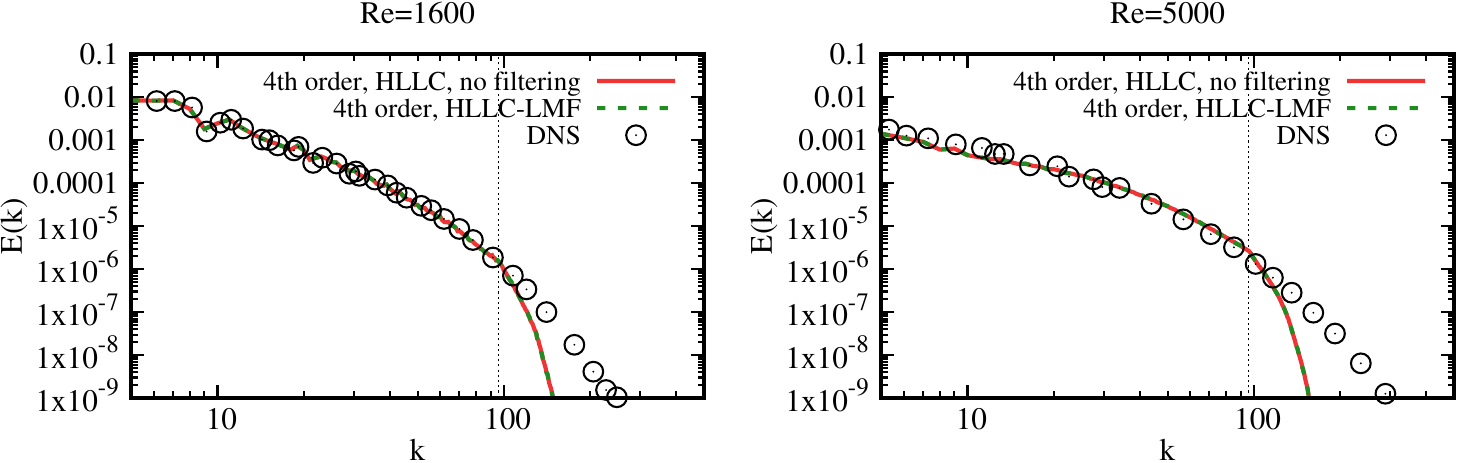}
\caption{Energy spectra of TGV at Re=1600 and 5000.}
\label{fig:tgv_k_spectral}       
\end{figure}

The energy spectra is also examined at both Reynolds numbers and it is computed at $t=9$ and $t=14$, respectively. This is shown in Fig.~\ref{fig:tgv_k_spectral}. The spectra of the DNS data at Re=1600 is computed using a pseudo-spectral method on a $512^3$ mesh by Carton de Wiart et al.~\cite{CartondeWiart2013} and the one for Re=5000 is again computed by Dairay et al.~\cite{Dairay2017}. The maximum resolvable wavenumber is marked by a vertical dash line. It can be seen that there is excellent agreement with the DNS data and there is no noticeable difference when LMF is used or not. It is noted that the purpose of this study is to demonstrate that when the flow is smooth, LMF essentially has a negligible effect on the flow field. But for more complicated flows, LMF will be switched on and this will be demonstrated in the following two turbine cases.  

Finally, in terms of computational cost, the computation with LMF is approximately $5.5\%$ more expensive than the one without filtering. This is the overhead introduced by computing the smoothness indicator $s_e$ and comparing them with the smoothness criterion $s_0$.  

\subsection{T106c Low Pressure Turbine}
This test case performs the spanwise periodic ILES of the transitional and separated flow on the T106c turbine cascade. This test case has been well studied in the literature and is a standard test case in the International Workshop on High-Order CFD Methods. The flow in the T106c turbine is subsonic and is calculated at Re = 80,000. The Reynolds number is based on the blade chord and the flow conditions at the exit boundary. The inlet turbulence is very low and the flow features laminar separation and transition in the reattachment zone. The details of the geometry and boundary conditions are summarized in Table~\ref{tab::t106c_geo}.

\begin{table}
\centering
\caption{Configurations of T106c turbine}
\label{tab::t106c_geo}
\begin{tabular}{ c c }
\hline \hline
 Item & Value \\ \hline
 s/c & 0.95  \\ 
 chord (c) & 1 \\
 Inlet flow angle & $32.7^{\circ}$  \\  
 Inlet total pressure & 669.30 Pa \\  
 Inlet total temperature & 298.15 K \\    
 Exit pressure & 503.87 Pa \\    
 Exit Mach number & 0.65 \\ 
 Freestream turbulence & $0.0 \%$ \\ 
 Exit Re & 80000 \\ \hline \hline
\end{tabular}
\end{table}

\begin{figure}
\centering 
\includegraphics[height=7cm]{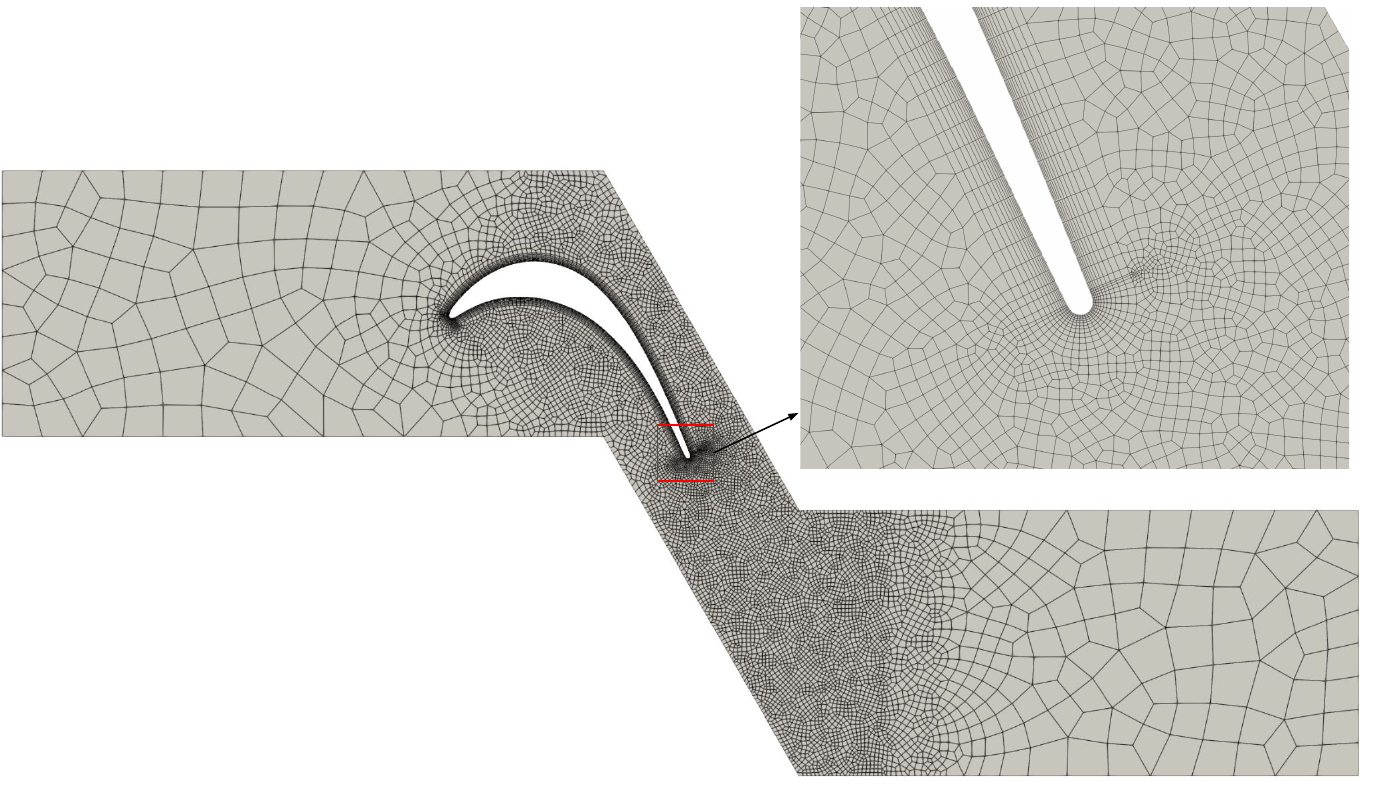}
\caption{Mesh details of the T106c turbine case.}
\label{fig:t106c_mesh_geo}       
\end{figure}

The grid is generated by Gmsh~\cite{Geuzaine2009} and the script is provided by the 4th International Workshop on High-Order CFD Methods~\footnote{https://how4.cenaero.be/content/as2-spanwise-periodic-dnsles-transitional-turbine-cascades}. A quadrilateral mesh is generated in the blade-to-blade section and a hexahedral mesh is generated by extruding the 2D quadrilateral mesh by $10\%$ of the chord for 12 layers. The script for the ``baseline" mesh from the workshop is used and the only modification to the original script is that the expansion ratio of the boundary layer mesh is reduced from 1.3 to 1.2. The total number of hexahedral cells is 122904. For a $p3$ simulation, the total number of DOFs per equation is 7865856. The axial distance between the leading edge and the inlet boundary is approximately $1.6c$, where $c$ is the blade chord. The axial distance between the trailing edge and the exit boundary is approximately $2.4c$. Figure~\ref{fig:t106c_mesh_geo} shows the quadrilateral mesh in the blade-to-blade section and a close-up view of the boundary layer mesh around the trailing edge.

\begin{table}
\centering
\caption{Numerical setup of T106c turbine case}
\label{tab::t106c_setup}
\begin{tabular}{cccccc}
\hline \hline
 Setup & order & $s_0$ & Riemann Solver & Filter & Success \\ \hline
 HLLC-LMF & p3 & -4.0 & HLLC &  LMF & \ding{51}  \\ 
 Roe-LMF & p3 & -4.0 & Roe & LMF & \ding{51}    \\ \hline \hline
  \end{tabular}
\end{table}

Table~\ref{tab::t106c_setup} shows the numerical setup for the T106c turbine. LMF is used to stabilize the computation, HLLC and Roe Riemann solvers are used and compared. The values of $\alpha$ and $s$ are set to 36 and 6, respectively, and their values are fixed throughout this paper. According to Hesthaven and Warburton~\cite{Hesthaven2007-yn} both values provide a reasonable balance of numerical accuracy and stability. $s_0$ is the flow smoothness criterion and determines how "local" the filter will be. If a large negative number is used, say $-16$, the filter will be applied to all elements. A value of -4 is used for both HLLC and Roe and successfully stabilizes the computation. More detailed analysis of the smooth criterion can be found in Section~\ref{sec:smooth_cri_analysis}.  

The computation starts from a $p_0$ run to efficiently wash away the initial transient flow field and creates a suitable initial flow field. Then the computation is restarted and the polynomial order is increased to $p1$. The same procedures are then used to increase the order of accuracy from $p1$ to $p2$, and eventually from $p2$ to $p3$. A fixed CFL number of 0.85 is used in the calculation. For different spatial orders of accuracy, this value is scaled by $\frac{1}{2p+1}$, where $p$ is the order of polynomial. For a $p3$ computation, the average time step is approximately $0.47 \times 10^{-4} t_c$ for HLLC-LMF,
where $t_c$ is the characteristic time and is defined as:
\begin{equation}
    t_c = \frac{c}{u_{ex}}
\end{equation}
where $c$ is the chord and $u_{ex}$ is the magnitude of the flow velocity on the exit boundary. The time step for Roe-LMF is very similar to HLLC-LMF. 

\begin{figure}
\centering 
\includegraphics[height=7.5cm]{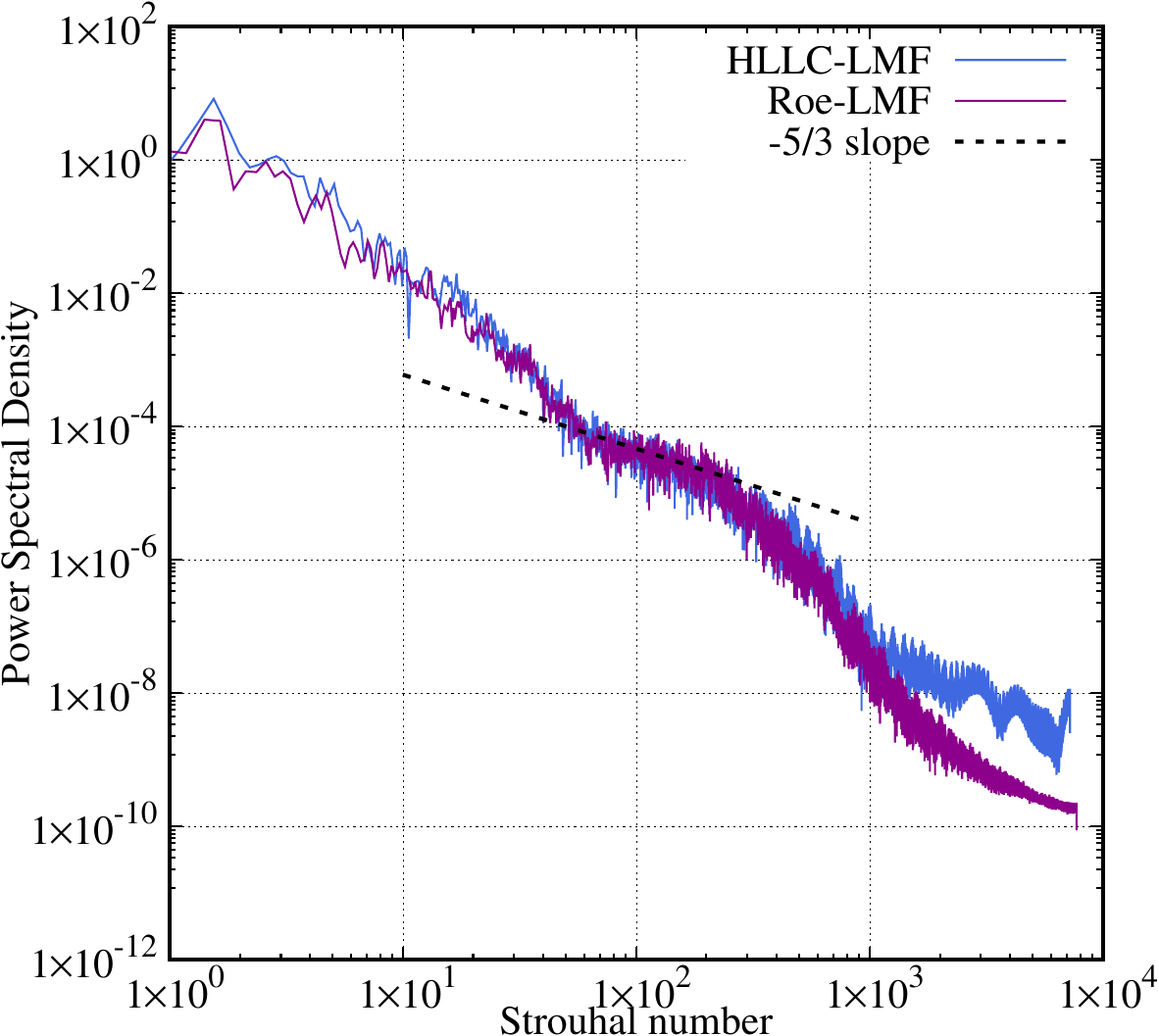}
\caption{Power spectral density of pressure for different numerical setups of the T106c turbine.}
\label{fig:t106c_psd}       
\end{figure}

Figure~\ref{fig:t106c_qcri_v} shows the isosurfaces of the Q criterion for the instantaneous flow calculated by HLLC-LMF. Transitional flows on the suction side and vortex shedding can be clearly seen. Figure~\ref{fig:t106c_psd} shows the power spectral density (PSD) of the pressure at a point downstream of the trailing edge. The point is on the middle span plane and its coordinate on the blade-to-blade section is $(0.8591 -0.5137)$. The frequency is represented in terms of the Strouhal number as:

\begin{figure}
\centering 
\includegraphics[height=7.5cm]{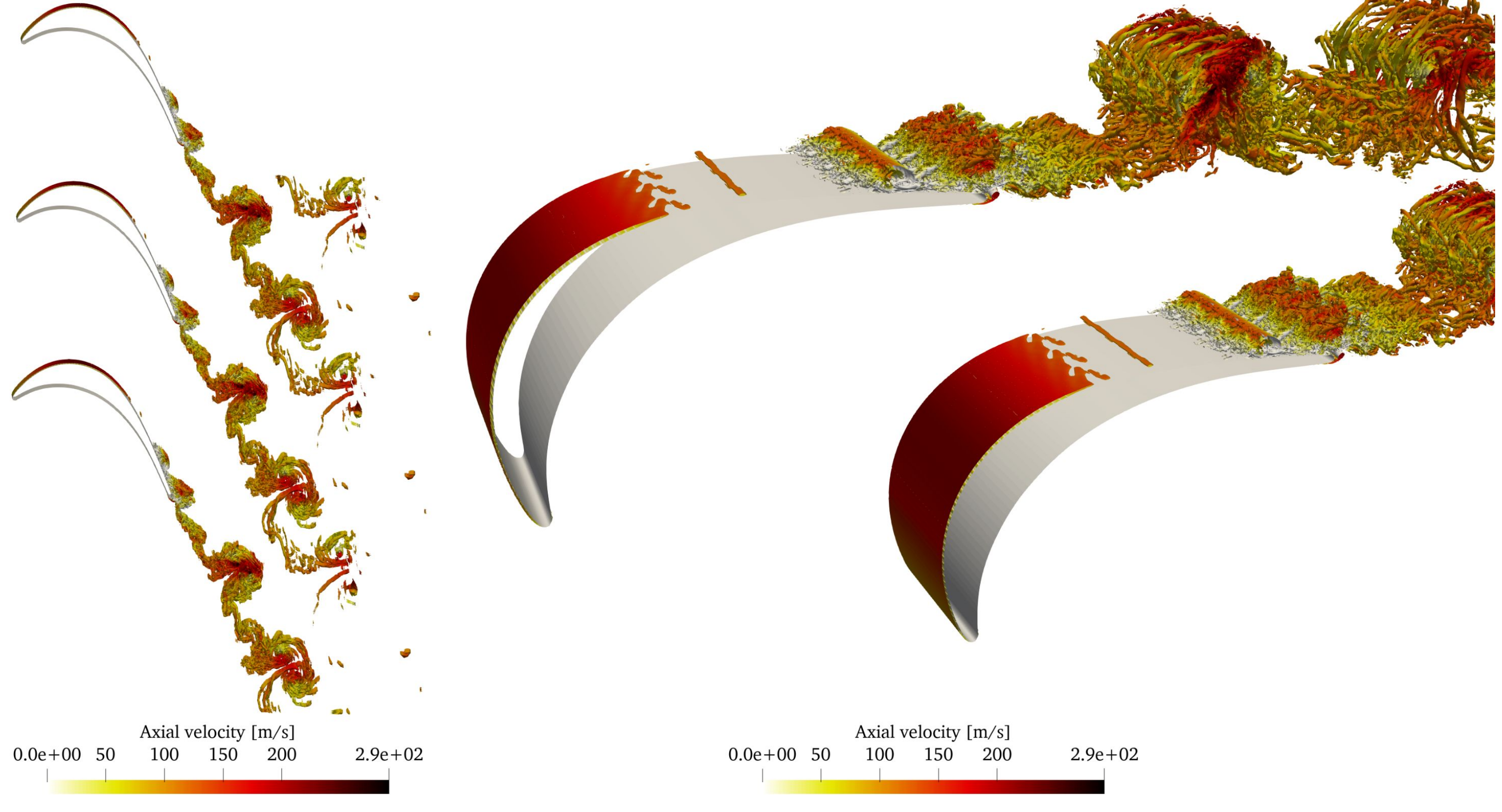}
\caption{Q-criterion of the instantaneous flow of the T106C turbine using the HLLC-LMF setup. The iso-surface is colored by axial velocity. The flow field has been replicated for three blade passages and then three times in the span-wise direction .}
\label{fig:t106c_qcri_v}       
\end{figure}

\begin{equation}
    \text{St} = \frac{f L_{char}}{u_{char}}
\end{equation}
in which $f$ is the frequency, $L_{char}$ is the characteristic length and $u_{char}$. In this T106c case, $L_{char}$ is the chord $c$ and $u_{char}$ is the velocity magnitude at the exit boundary. For the time-step that is used in the computation, the cut-off Strouhal number $St_{\text{cutoff}} \approx 6300$. Figure~\ref{fig:t106c_psd} shows that for this probed point both Roe and HLLC agree with each other in the inertial sub-range, but there are noticeable differences at high frequencies. The reason for such differences can be related to the inherent numerical dissipation in both Riemann solvers and this would require further investigations.

\begin{figure}
\centering 
\includegraphics[height=3cm]{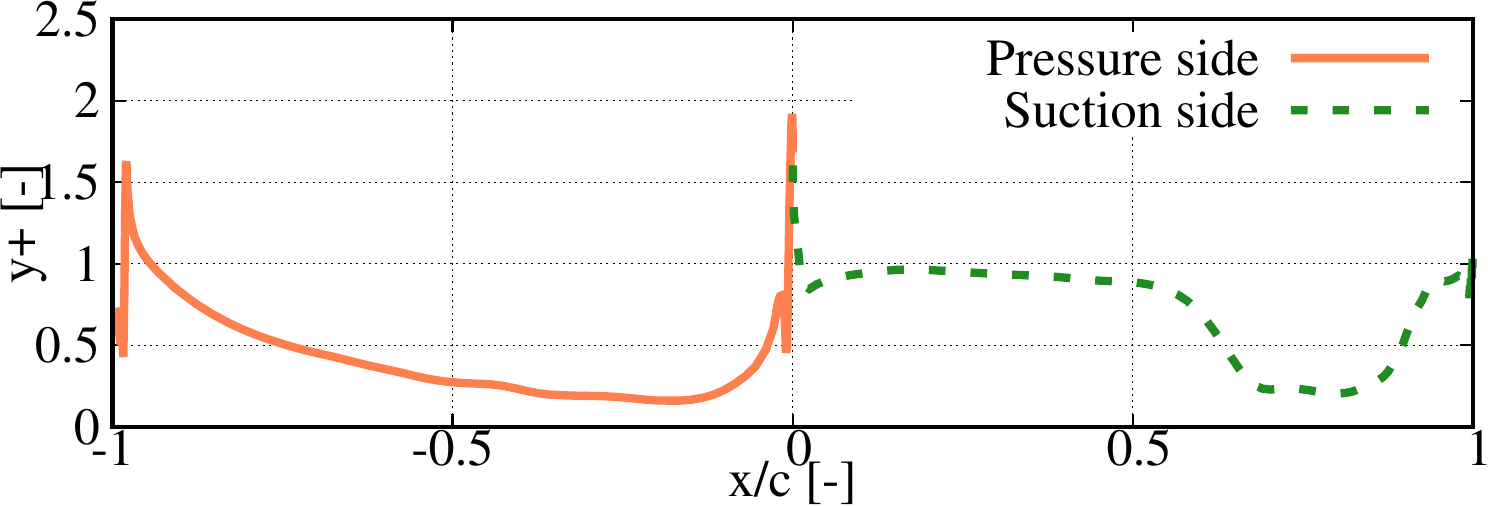}
\caption{Mean flow $y^+$ of the T106c turbine case using HLLC-LMF.}
\label{fig:t106c_yplus}       
\end{figure}

The mean flow is computed by averaging the snapshots of the instantaneous flows. Witherden et al.~\cite{Witherden2016} obtained the mean flow by averaging the flows for 2$t_c$ and the results were in good agreement with the experimental data. In this work, 5$t_c$ is used to calculate the averaged flow and this should be sufficient to obtain the mean flow. Figure~\ref{fig:t106c_yplus} shows the $y^+$ of the first cell on the blade, and it is computed as:
\begin{equation}
    y^+ = \frac{\rho u_{\tau} d}{\mu} \frac{1}{p+1}
\end{equation}
in which $u_{\tau}$ is the local friction velocity of the time-averaged flow and $d$ is the height of the first cell layer in the wall normal direction. The term $\frac{\rho u_{\tau} d}{\mu}$ would be  the $y^+$ for a finite volume code. However, for FR or DG,  as the number of DOFs in the normal direction of the wall varies with the order of the polynomial, the calculated $y^+$ using the cell height must be scaled by the order of spatial accuracy $p+1$~\cite{Jia2019} to reflect this. It can be seen that the averaged $y^+$ on the blade surface for $p3$ is approximately 0.5. In terms of $x^+$ and $z^+$, which represent streamwise and spanwise near-wall grid spacing, their averaged values on the blade surface are approximately 20 and 6, respectively.



\begin{figure}
\centering 
\includegraphics[height=5cm]{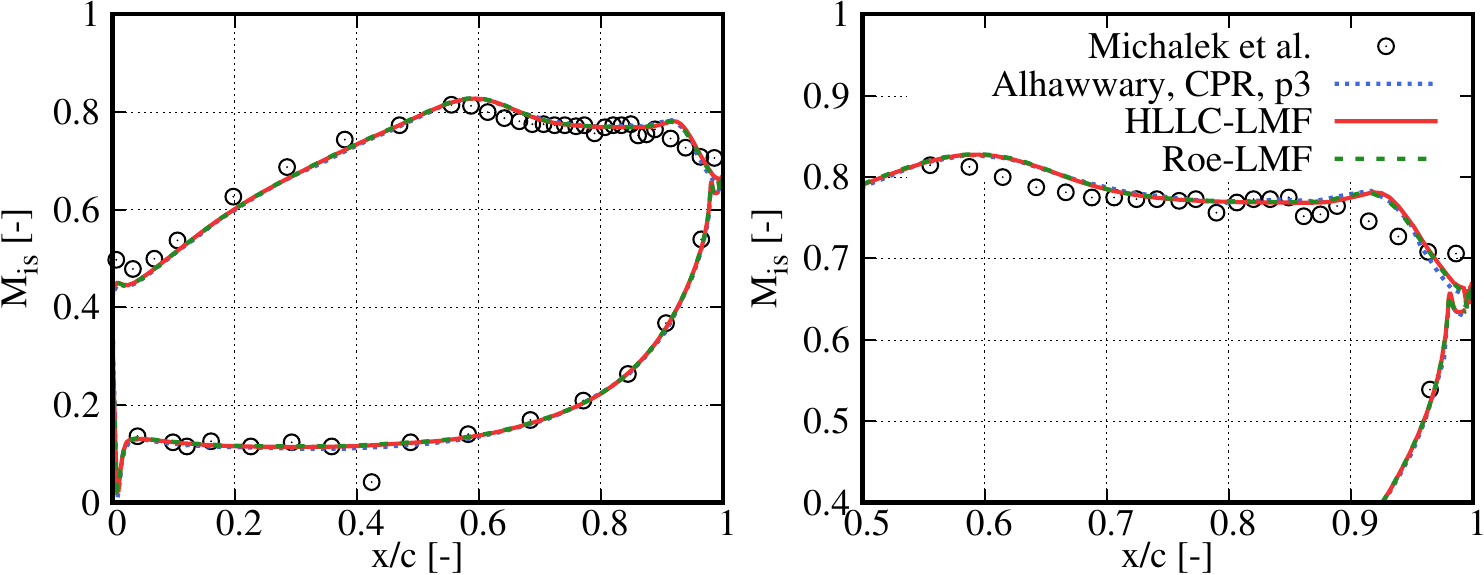}
\caption{Predicted isentropic Mach number of the T106c turbine.}
\label{fig:t106c_mach}       
\end{figure}

Figure~\ref{fig:t106c_mach} shows the predicted isentropic Mach number against the experimental data of Michalek et al.~\cite{Michlek2012}. The solution of Alhawwary and Wang~\cite{Alhawwary2019} is used as a reference and calculated using the CPR method, and the polynomial order is $p3$. From the comparison, it can be seen that the solutions of Roe-LMF and HLLC-LMF have excellent agreement with the experimental data. Furthermore, the solutions of HLLC-LMF, Roe-LMF, and CPR-$p3$ are very similar to each other for the laminar and transition flow regions. This shows that in this case applying LMF has a negligible impact on the prediction of the isentropic Mach number. 

\subsection{High Pressure Turbine Vane with Heat Transfer}
The third test case is the spanwise periodic ILES simulation of the linear cascade of a high-pressure turbine vane LS89. The Mach number, Reynolds number, and temperature ratio between the inlet and exit are representative of typical industrial high-pressure turbines. The experimental data are available from Arts et al.~\cite{arts_report}. The experiment involves studies of an uncooled turbine vane under a range of the Mach number, Reynolds number, and freestream turbulence. These conditions provide challenging and well-suited test cases for predicting the transition and heat transfer on the blade surface. There have been several successful LES/DNS simulations of this case using the high-order finite-difference~\cite{Wheeler2016} method,  finite-volume approach\cite{Dupuy2020}, and CPR~\cite{Jia2019}. The flow condition of the current study is MUR129. This case has a very low level of incoming turbulence ($0.8 \%$) at the inlet. Previous work~\cite{Jia2019} prescribes no freestream turbulence at the inlet and obtains excellent agreement with the experimental data. The geometry detail and the flow condition for MUR129 that are used in this study are summarized in Table~\ref{tab::ls89_detail}. The heat transfer coefficient (HTC) and the isentropic Mach number will be used to validate the current FR solver against the experimental data and to demonstrate the impact of different numerical configurations.

\begin{table}
\centering
\caption{Configurations of VKI LS89 at the flow condition MUR129}
\label{tab::ls89_detail}
\begin{tabular}{l l }
\hline \hline
 Item & Value \\ \hline
 chord (c) & 0.0676 m  \\ 
 pitch & 0.05749 m  \\  
 span length & 0.0112 m \\  \hline
 Inlet total pressure & 1.849 bar \\    
 Inlet total temperature & 409.2 K \\    
 Inlet flow angle & 0.0 \\ 
 Freestream turbulence & $0.0 \%$ \\ 
 Exit pressure & 1.82 bar \\ 
 Exit Re & $1.13 \times 10^6$ \\ 
 Wall temperature & 297.75 K \\ \hline \hline
\end{tabular}
\end{table}

\begin{figure}
\centering 
\includegraphics[height=9cm]{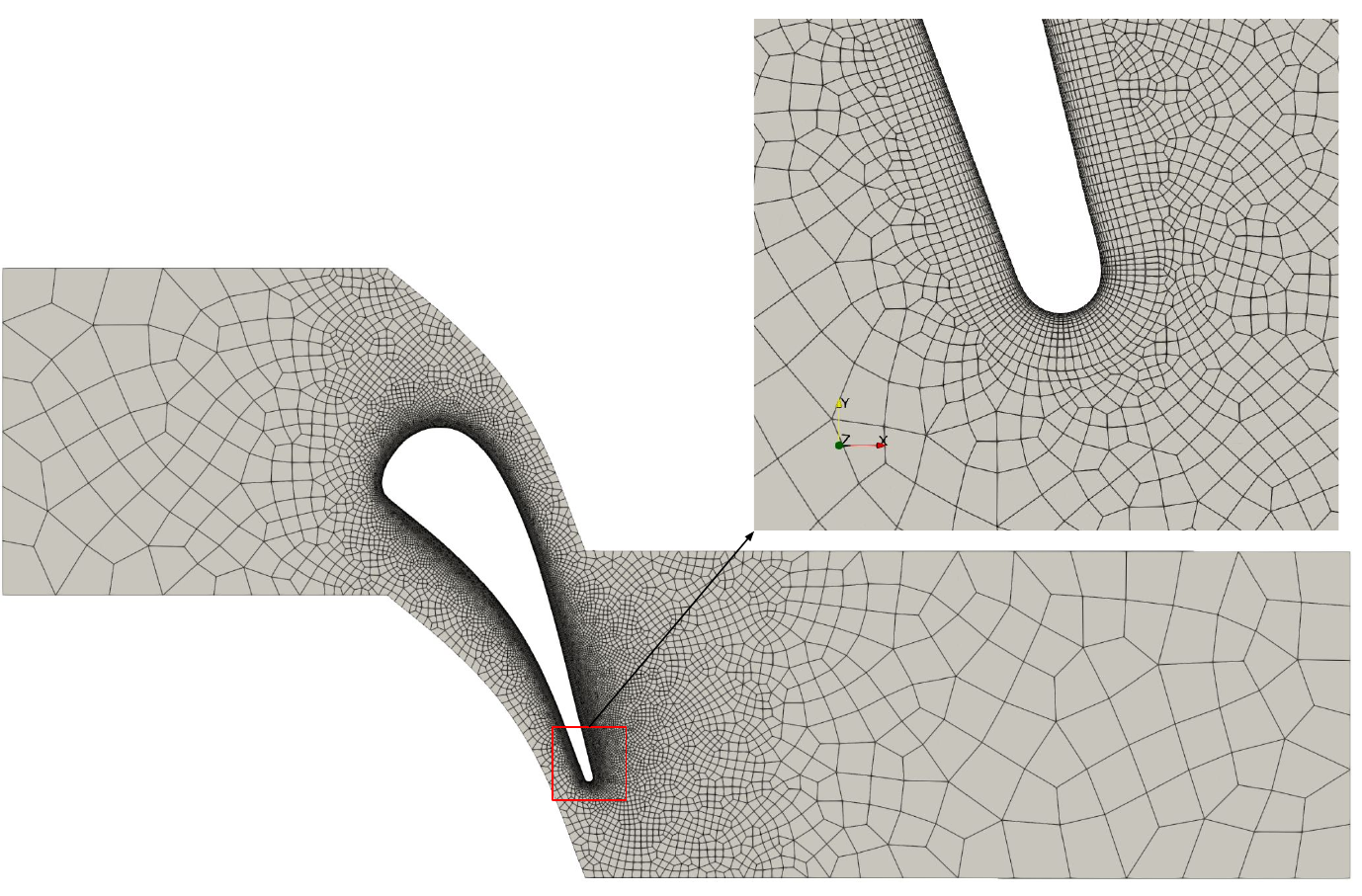}
\caption{Mesh details of the VKI LS89 turbine.}
\label{fig:ls89_mesh}       
\end{figure}

The grid is generated by Gmsh. A quadrilateral mesh is generated in the blade-to-blade section and then extruded in the spanwise direction by $16.6\%$ of the chord. This value was used in the previous work by Jia and Wang~\cite{Jia2019} and shows a good prediction of HTC on the blade surface. Furthermore, Morata et al.~\cite{ColladoMorata2012} shows that the maximum difference of HTC on the blade surface is within $5\%$ when the span extrusion increases from $10\%$ to $20\%$ of the chord, but when the extrusion is reduced to $5\%$ of the chord, obvious differences in HTC are observed. Therefore, extrusion of the $16.6\%$ chord is a cost-effective option.   Figure~\ref{fig:ls89_mesh} shows the unstructured quadrilateral mesh in the blade-to-blade section and a close-up view of the mesh around the trailing edge. Several levels of spanwise grid spacing will be studied. In the coarsest one, 10 layers of hexahedra are extruded in the spanwise direction, and the total number of hexahedra is 244150. For the $p3$ simulation, the total number of degrees of freedom (DOF) per equation is 15625600.  The axial distance between the leading edge and the inlet boundary is approximately $c$. The axial distance between the trailing edge and the exit boundary is approximately $2c$. These distances are slightly larger than previous studies~\cite{Jia2019,Dupuy2020} and this is beneficial to reduce the impact of the inflow/outflow boundaries on the solution. At the inlet, the total pressure, the total temperature and two flow angles are specified. At the exit, a static pressure is specified. The upper and lower surfaces are set to be periodic.

The computation starts with a $p_0$ run that efficiently creates a suitable initial flow field. Then the computation is restarted and the spatial precision order is increased to $p1$. The same procedures are then used to increase the order of accuracy from $p1$ to $p2$, and eventually from $p2$ to $p3$. A fixed CFL number of 0.85 is used in the computation. For different spatial order of accuracy, this value is scaled by $\frac{1}{2p+1}$, where $p$ is the order of polynomials.  For a $p3$ calculation, the average time step for HLLC-LMF is approximately $0.56 \times 10^{-5} t_c$, which $t_c$ is the characteristic time. 

Regarding the mean flow, it is computed by averaging the flow for 5$t_c$, where $t_c$ is the characteristic time and is defined as:
\begin{equation}
    t_c = \frac{c}{u_{ex}}
\end{equation}
where $c$ is the chord and $u_{ex}$ is the magnitude of the flow velocity on the exit boundary. 

\subsubsection{Effect of DOFs}

For a typical second-order finite-volume solver, DOFs in the computation are determined by the grid size. However, for a spectral element solver (i.e. FR), both the grid size ($h$) and the polynomial order ($p$) can be adjusted to increase the DOFs in the computation. It is not computationally feasible to perform a sensitivity study of both the grid size (i.e. in streamwise, wall normal, and spanwise directions) and the polynomial order altogether for the LES of an industrial case. In this work, a segmented approach is used. In the first place, a series of 2D calculations have been performed to determine a desirable grid spacing in the streamwise ($x^+$) and the normal direction ($y^+$) on the blade surface for a $p3$ calculation. For simplicity, this 2D study is not shown here. The resulting 2D mesh is then used to perform the sensitivity study in 3D. This study is summarized in Table~\ref{tab::ls89_grid_sensi}. Mean flow quantities that are relevant to engineering designs, such as skin friction, the isentropic Mach number ($\text{M}_{\text{is}}$) and HTC are used as criteria to demonstrate the effect of increasing the polynomial order $p$ and reducing the grid size $h$ on the solution, respectively.

\begin{table}
\centering
\caption{Sensitivity study of grid size and polynomial order for the LS89 turbine case}
\label{tab::ls89_grid_sensi}
\begin{tabular}{ccccccc}
\hline \hline
 Setup & order & $s_0$ & Rie. Solver & Filter & No. of hex Layer &DOFs \\ \hline
 1 & p2 & -3.3 & HLLC & LMF  & 10 & 6592050    \\
 2 & p3 & -4.0 & HLLC & LMF  & 10 & 15625600  \\
 3 & p3 & -4.0 & HLLC & LMF  & 15 & 23438400  \\
 4 & p3 & -4.0 & HLLC & LMF  & 20 & 31251200  \\ \hline \hline
 \end{tabular}
\end{table}

\begin{figure}
\centering 
\includegraphics[height=8cm]{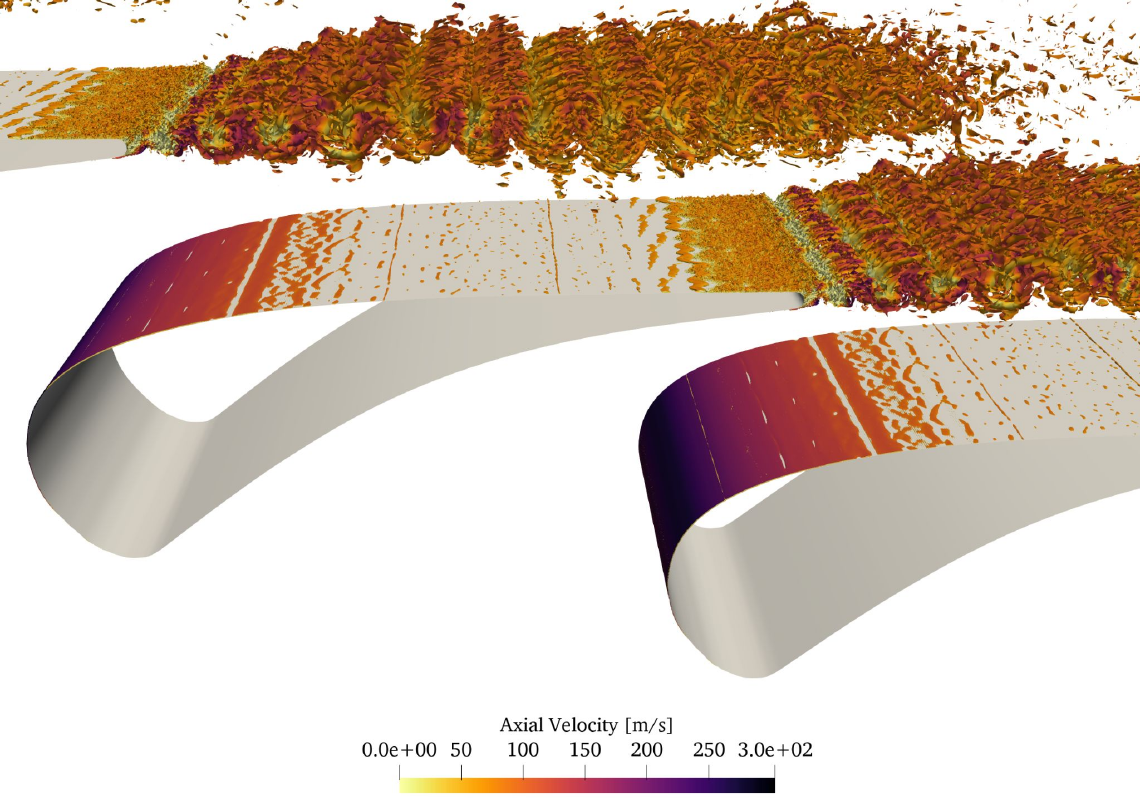}
\caption{Q-criterion of the instantaneous flow of the LS89 turbine using the HLLC-LMF setup. The iso-surface is colored by axial velocity. The flow field has been replicated for three blade passages and then three times in the span-wise direction.}
\label{fig:ls89_qcri}       
\end{figure}

\begin{figure}
\centering 
\includegraphics[height=4cm]{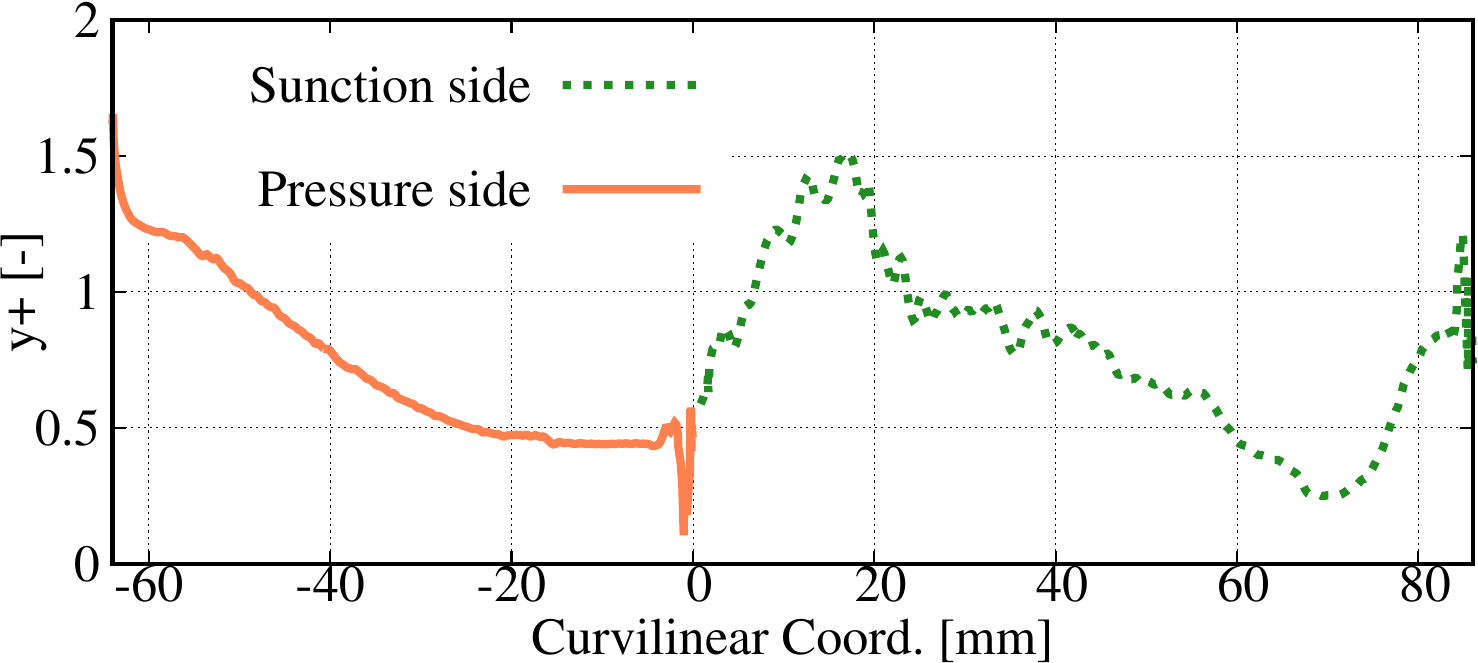}
\caption{Typical $y^+$ on the blade surface of a p3 computation of the the LS89 turbine.}
\label{fig:ls89_yplus}       
\end{figure}

The HLLC Riemann solver is used in this sensitivity study and the choice of $s_0$ for LMF is guided by Table~\ref{tab::smooth_indicator_value_new}. $p2$ and $p3$ computations are performed on a mesh with 10 hexahedral layers in the spanwise direction. This is then increased to 15 and 20 layers. Figure~\ref{fig:ls89_qcri} shows the isosurface of the instantaneous flow Q criterion of Setup 4, and the isosurface is colored by the axial velocity. The flow is laminar on the pressure side but transitional on the suction side and then becomes fully turbulent towards the trailing edge. This flow structure is representative for all the setups.  Figure~\ref{fig:ls89_yplus} shows the representative $y^+$ on the blade for Setups 2,3 and 4. It can be seen that the averaged $y^+$ on the blade surface for $p3$ is approximately 0.75. Regarding $x^+$, which represents the grid spacing in the streamwise, they are approximately 30 for Setups 1-4. With respect to $z^+$, they are on average 90, 60 and 45 on the blade surface for Setups 1$\&$2, 3 and 4, respectively. 

\begin{figure}
\centering 
\includegraphics[height=8cm]{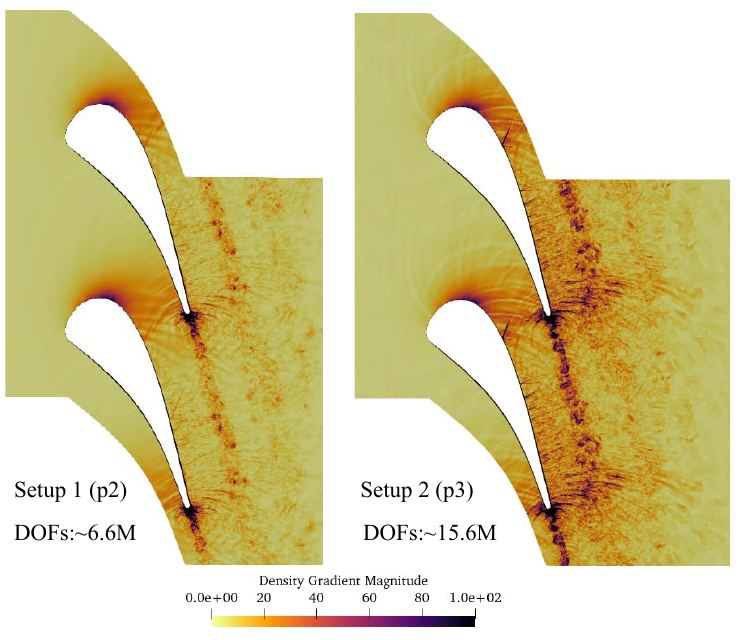}
\caption{Magnitude of density gradient for $p2$ and $p3$ of the VKI LS89 turbine.}
\label{fig:ls89_p2_p3}       
\end{figure}

Figure~\ref{fig:ls89_p2_p3} shows the comparison of the instantaneous flow density gradient for setup 1 and setup 2 in Table~\ref{tab::ls89_grid_sensi}. Qualitatively, it can be seen that the $p3$ computations produce a higher resolution of the wakes and pressure waves that are radiated from the trailing edge.

\begin{figure}
\centering 
\includegraphics[height=9cm]{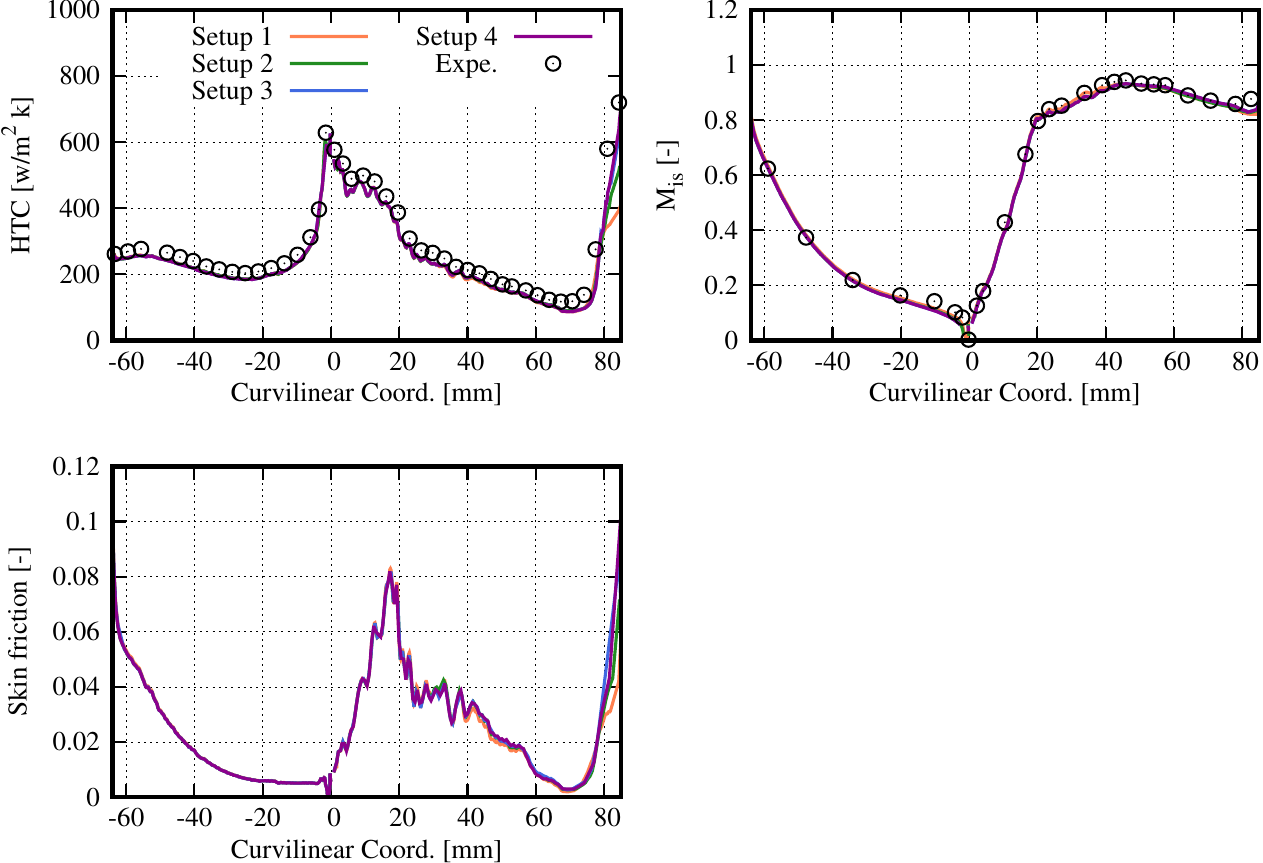}
\caption{Sensitivity study of HTC, isentropic Mach number and Skin friction on DOFs. }
\label{fig:ls89_grid_sensi}       
\end{figure}

Figure~\ref{fig:ls89_grid_sensi} shows the comparisons of $\text{M}_{\text{is}}$, HTC and skin friction for these 4 setups in Table~\ref{tab::ls89_grid_sensi}. The data are plotted on the basis of the curvilinear coordinates on the blade surface starting from the leading edge. Coordinates with positive values are on the suction side and those with negative values are on the pressure side.

A clear convergence can be observed when the DOFs are increased from setup 1 to setup 4 and a better agreement with the experimental data is also observed. Several observations can be made: 
\begin{itemize}
    \item $\text{M}_{\text{is}}$ is not sensitive to the variation of DOFs in 3D. In fact, the report of Arts~\cite{arts_report} showed that an inviscid 2D computation can already obtain a reasonably good prediction of $\text{M}_{\text{is}}$. This is because the flow is mostly laminar on the blade surface, the turbulent flow patch toward the trailing edge on the suction side has only a marginal effect on $\text{M}_{\text{is}}$. This also shows that only comparing $\text{M}_{\text{is}}$ with the experimental data for the LS89 turbine is not sufficient to validate the CFD solver. 
    \item All these setups have predicted the location of the laminar-to-turbulent transition well compared to the experimental data. For the compared flow quantities, there is no noticeable difference in the regions where the flow is laminar. 
    \item An obvious difference is observed in the turbulent-flow region. The skin friction and HTC has increased significantly due to enhanced momentum and energy transfer from the turbulent flow. And an improvement of the prediction of HTC can be observed when DOFs are increased from Setup 1 to Setup 4.
\end{itemize}

\begin{figure}
\centering 
\includegraphics[height=5cm]{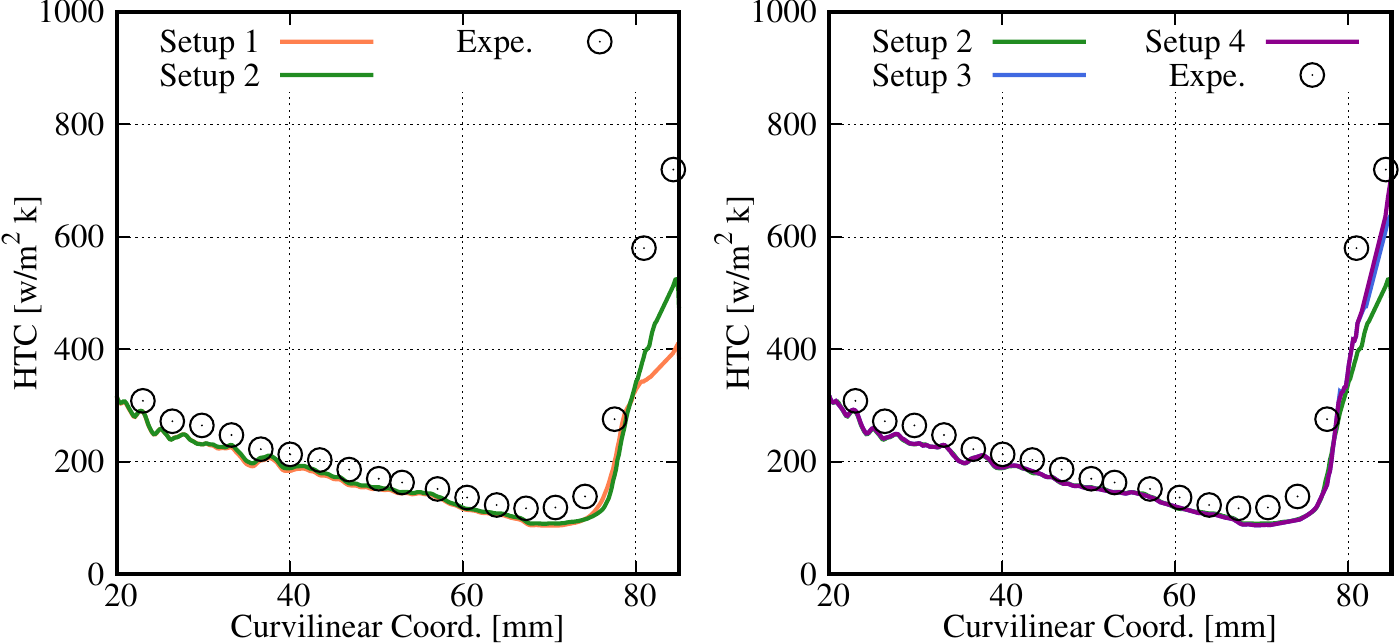}
\caption{Sensitivity study of HTC on polynomial order $p$ and grid size $h$.}
\label{fig:ls89_p_effect}       
\end{figure}

Figure~\ref{fig:ls89_p_effect} shows a more detailed comparison of the effect of DOFs on HTC predictions. The figure on the left shows the impact of the polynomial order $p$ on HTC prediction. It shows that with the same grid size when the polynomial order increases from 2 to 3 (the DOFs increase by a factor of $\frac{4^3}{3^3}\approx2.4$), the HTC prediction has been improved towards the trailing edge on the suction side. The figure on the right shows that with the same polynomial order $p$, reducing the gird spacing $h$ in the spanwise direction further improves the HTC prediction.  As is mentioned before, Setups 2, 3 and 4 have 10, 15 and 20 layers in the spanwise direction, respectively. Figure~\ref{fig:ls89_p_effect} shows that for $p3$ calculations using 15 layers is a cost-effective option to obtain a good HTC prediction, and this corresponds to an average $z^+$ of 60 on the blade surface. 

In turbomachinery simulations, because the flow gradient in the streamwise direction is larger than that in the spanwise direction, it is common for the grid spacing in the span-wise direction to be coarser than the one in the streamwise direction to reduce computational cost. Alhawwary and Wang~\cite{Alhawwary2019} showed that for a low-pressure turbine, coarsening the grid in the spanwise direction is less critical in predicting the mean flow quantities. The results in Figure~\ref{fig:ls89_p_effect} show that this statement also holds for the HTC prediction. 
 

\subsubsection{Effect of Riemann Solvers and Filters}
 
Spiegel et al.~\cite{Spiegel2015} show that aliasing-driven numerical instability is more likely to occur on a coarse grid due to undersolved flow features. To demonstrate the effect of Riemann solvers and filters on the numerical stability of the FR computation, Setup 2 in Table~\ref{tab::ls89_grid_sensi} is selected, as it is the $p3$ run with the fewest DOFs, but still produces satisfactory results compared to the experimental data. Table~\ref{tab::ls89_setup} shows the different setups that are used to demonstrate the effect of the Riemann solvers and the filtering approach on the simulation of LS89. 

For the value of $s_0$ in LMF, -4 and -4.3 are attempted. HLLC can stabilize with both $s_0$ values, while Roe can only stabilize with $s_0 = -4.3$. This means that HLLC is more robust than Roe, as the computation can stabilize at a higher value of $s_0$. An explanation for this can be as follows. When a potential numerical instability develops, the density can approach zero and a local vacuum state could be formed. It is known~\cite{Toro2009-bg} that the Roe solver is not robust enough for this flow condition, while the HLLC is robust enough to handle it. With respect to EF, a second-order modal filter (see Equation~\ref{eqn::modal_filter_ef}) is used and the value of $\alpha$ is calculated adpatively for each element via an optimization process to satisfy the positivity and minimum entropy conditions.


\begin{table}
\centering
\caption{Different setups for the LS89 turbine simulation to study the effect of Riemann solvers and filter configurations}
\label{tab::ls89_setup}
\begin{tabular}{cccccc}
\hline \hline
 Setup & order & $s_0$ & Riemann Solver & Filter & Success \\ \hline
 HLLC-LMF & p3 & -4.3 & HLLC &  LMF & \ding{51}  \\ 
 Roe-LMF & p3 & -4.3 & Roe & LMF & \ding{51}    \\
 HLLC-LMF-2 & p3 & -4.0 & HLLC & LMF  & \ding{51}    \\
 Roe-LMF-2 & p3 & -4.0 & Roe & LMF  &  \ding{53}  \\
 HLLC-EF  & p3 & / & HLLC & EF & \ding{51}  \\ \hline \hline
 \end{tabular}
\end{table}

\begin{figure}
\centering 
\includegraphics[height=9.5cm]{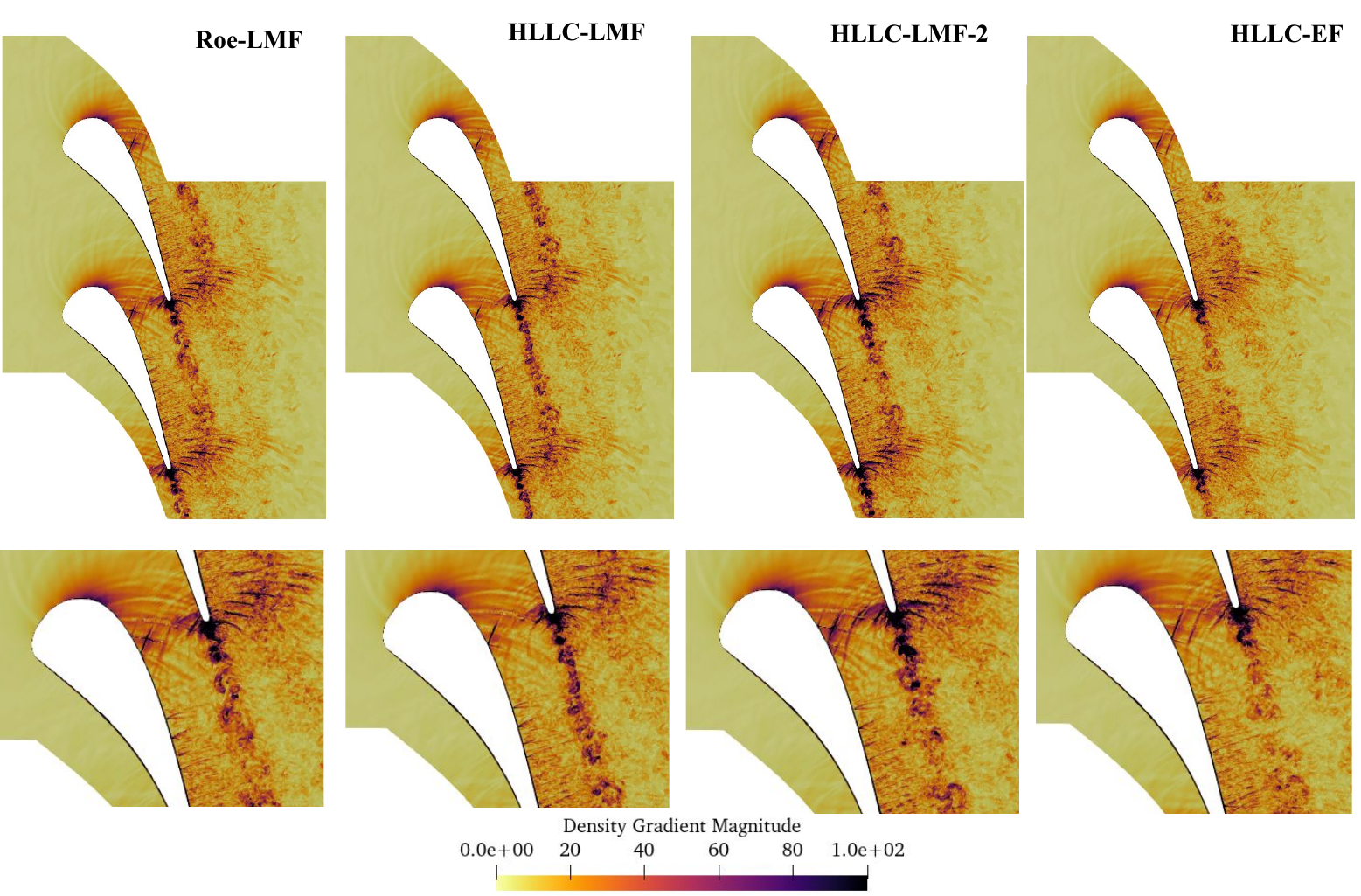}
\caption{Magnitude of density gradient for different filtering and Riemann solver configurations of the VKI LS89 turbine.}
\label{fig:ls89_grad_rho_vs}       
\end{figure}

The effect of the different setups in Table~\ref{tab::ls89_setup} on instantaneous flow can be qualitatively demonstrated by Fig~\ref{fig:ls89_grad_rho_vs}. The figure shows the magnitude of the density gradient in the middle-span section. All setups capture the dominant flow features of the unsteady flow, for example, the suction side transition, vortex shedding from the trailing edge, and the resulting pressure waves radiating from the trailing edge. HLLC-LMF shows a resolution of the pressure waves comparable to Roe-LMF. With a more relaxed smoothness criterion, HLLC-LMF-2 further improves the resolution of the pressure waves compared to HLLC-LMF and Roe-LMF. 

\begin{figure}
\centering 
\includegraphics[height=7cm]{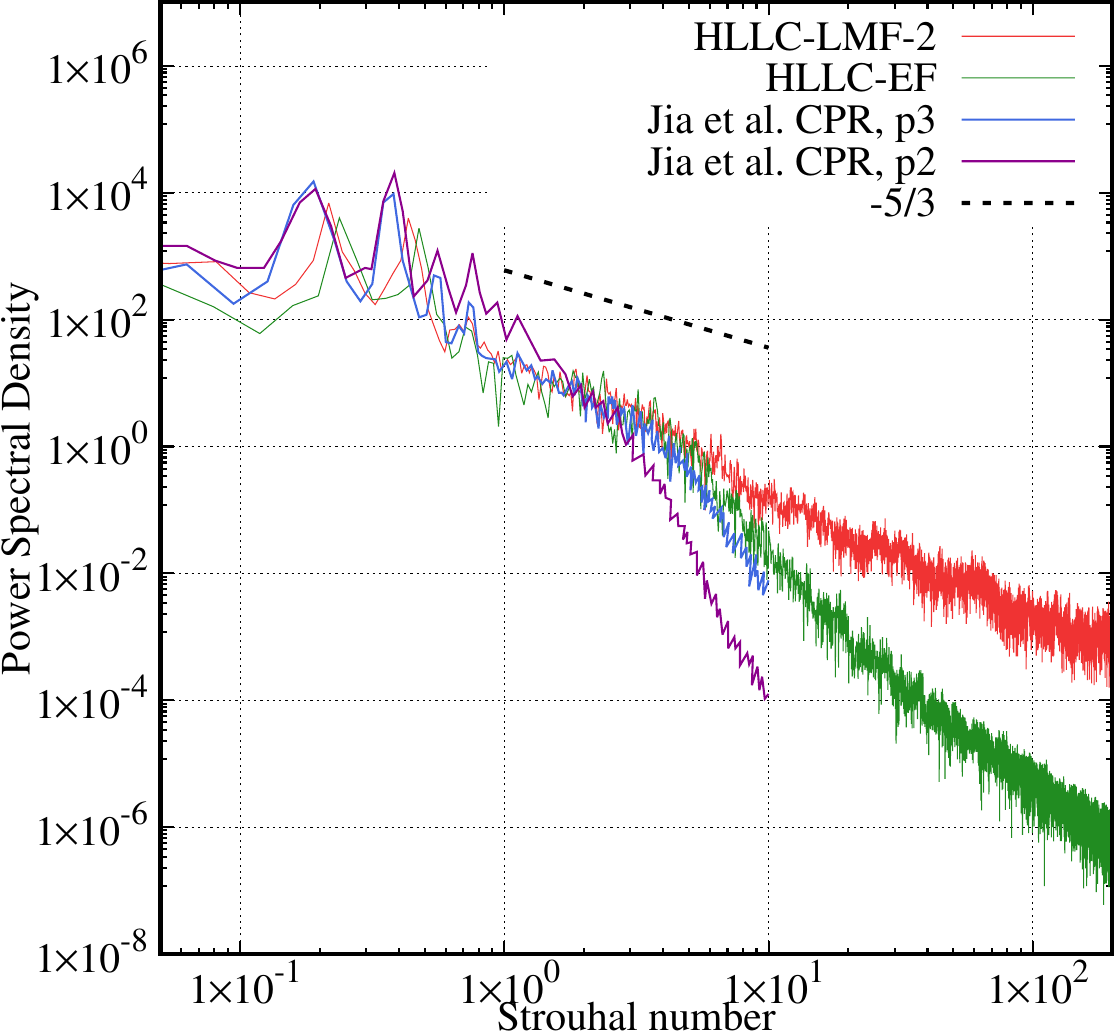}
\caption{Power spectral density of pressure for HLLC-EF and HLLC-LMF-2 of the LS89 turbine at the point $(0.03722, -0.05482, 0.0)$.}
\label{fig:psd}       
\end{figure}

With regard to EF, it successfully stabilizes the solution without any tuning parameters but shows a less sharp resolution of the pressure waves compared to HLLC-LMF-2. This observation is consistent with the results of Gao et al.~\cite{Gao2023}, who reported that EF could introduce slightly more dissipation into the computation. In order to demonstrate this, the PSD of pressure is monitored for a point, which is located downstream of the trailing edge on the mid-span plane. Its coordinate on the blade-to-blade section is $(0.03722, -0.05482)$.  The PSD for this coordinate was also computed by Jia et al.~\cite{Jia2019} and therefore their data is included. This comparison is shown in Fig.~\ref{fig:psd}. The frequency is represented in terms of the Strouhal number. The characteristic length $L_{char}$ is the radius of the trailing edge and the characteristic velocity $u_{char}$ is the magnitude of the velocity at the exit boundary. Based on the selected time step, the cutoff Strouhal number is $St_{\text{cut-off}} \approx 200$. For clarity, only the comparison of HLLC-LMF-2 and HLLC-EF is shown, and the PSD of other setups with LMF in Table~\ref{tab::ls89_setup} is found to be very similar to HLLC-LMF-2. The results of Jia are computed by the code hpMusic, which implements the CPR method. ILES with implicit time stepping is used, and the cut-off Strouhal number is around 10.

From Jia's CPR solutions, it can be seen that increasing the spatial accuracy from p2 to p3 shows a better agreement with the -5/3 rule on the inertial subrange. Jia's data also show that more numerical dissipation leads to a faster decay of PSD, especially at higher frequencies. When comparing the CPR p3 solution with HLLC-LMF-2, both data agree well on the inertial subrange, but the CPR p3 solution shows a faster decay of PSD at higher frequencies. This indicates that HLLC-LMF-2 is less dissipative than Jia's CPR p3 solution. Jia also compared the impact of explicit and implicit time stepping on PSD, the comparison showed that the choice of time-stepping did not have a noticeable impact on PSD. Therefore, the reason why Jia's CPR p3 solution is more dissipative than the HLLC-LMF-2 result could be caused by the limiter used in hpMusic to stabilize the computation. With respect to EF, it is interesting to observe that its PSD is similar to that of Jia's CPR p3 data. This shows that EF does introduce a slightly higher dissipation at higher frequencies than HLLC-LMF-2, but this does not seem to have a big impact on the inertial subrange. 


\begin{figure}
\centering 
\includegraphics[height=10cm]{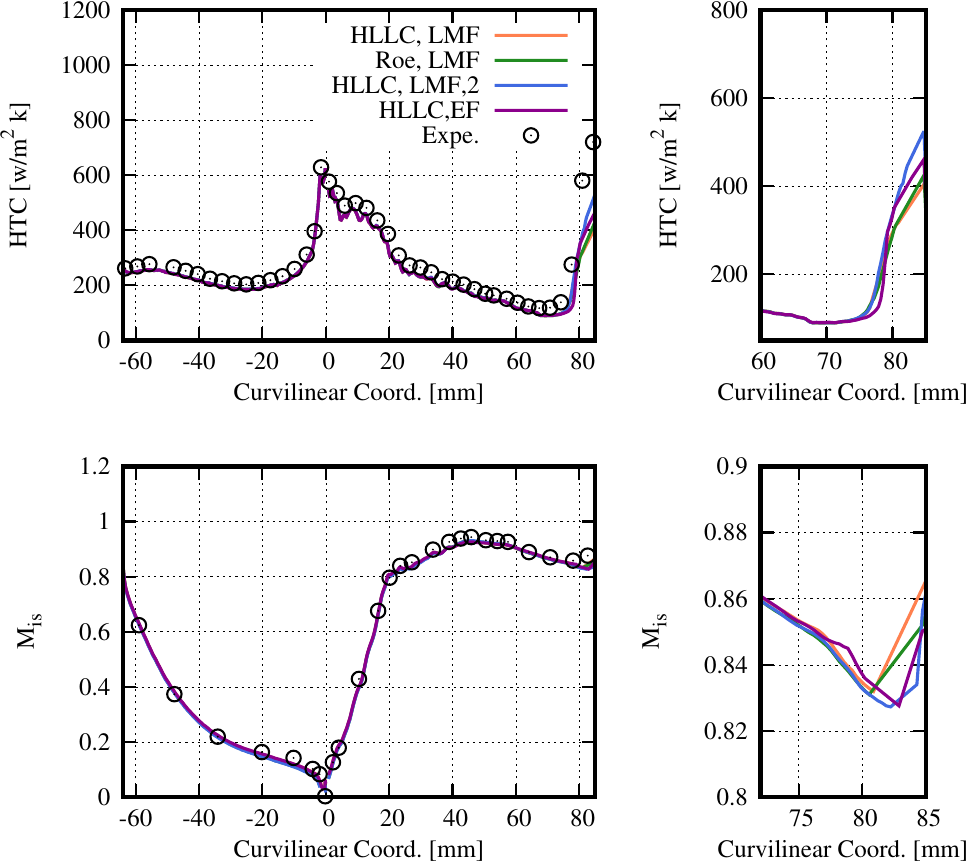}
\caption{HTCs and isentropic Mach numbers for different setups in Table~\ref{tab::ls89_setup} of the LS89 turbine.}
\label{fig:ls89_htc_mach}       
\end{figure}

Figure~\ref{fig:ls89_htc_mach} shows the predicted HTC and the isentropic Mach number of the time-averaged flow for the setups in Table~\ref{tab::ls89_setup}.  ``Roe-LMF" and ``HLLC-LMF" predicted similar HTCs, while "HLLC-LMF-2" shows an improved prediction.  This indicates that the predicted HTCs are more relevant to the filter than the choice of the Riemann solver. The differences of ``HLLC-LMF" and ``HLLC-LMF-2" also show that the prediction of the transition is less sensitive to $s_0$, but the predicted HTCs in the turbulent flow region are sensitive to $s_0$. EF does not require a tuning parameter for the filter and is successful in stabilizing the solution. HTC is slightly under-predicted compared to "HLLC-LMF-2" but shows a slight improvement over "HLLC-LMF" and "Roe-LMF". 

In terms of the isentropic Mach number, similar conclusions can be drawn as in Fig.~\ref{fig:ls89_p2_p3}: isentropic Mach number is insensitive to the configuration of filters and Riemann solvers in this case.


\subsection{Remarks on the filter and smoothness criterion}
Table~\ref{tab::smooth_indicator_value_new} shows the range of $s_0$ in which LMF could potentially be used to stabilize the computation and has been used as a guide to choose the values of $s_0$ in the T106c and LS89 turbine cases. Here, we perform a posteriori analysis of the smooth criterion using the values of $s_e$ for the T106c and LS89 turbine cases. To the best knowledge of the author, there has been no previous work showing the variation of flow smoothness $s_e$ for turbomachinery flows. Therefore, it is useful to perform such an analysis as it can provide a rationale for the range of $s_0$ in Table~\ref{tab::smooth_indicator_value_new} and guide the selection of a suitable $s_0$.
\subsubsection{The T106c Turbine}
\label{sec:smooth_cri_analysis}

\begin{figure}
\centering 
\includegraphics[height=10cm]{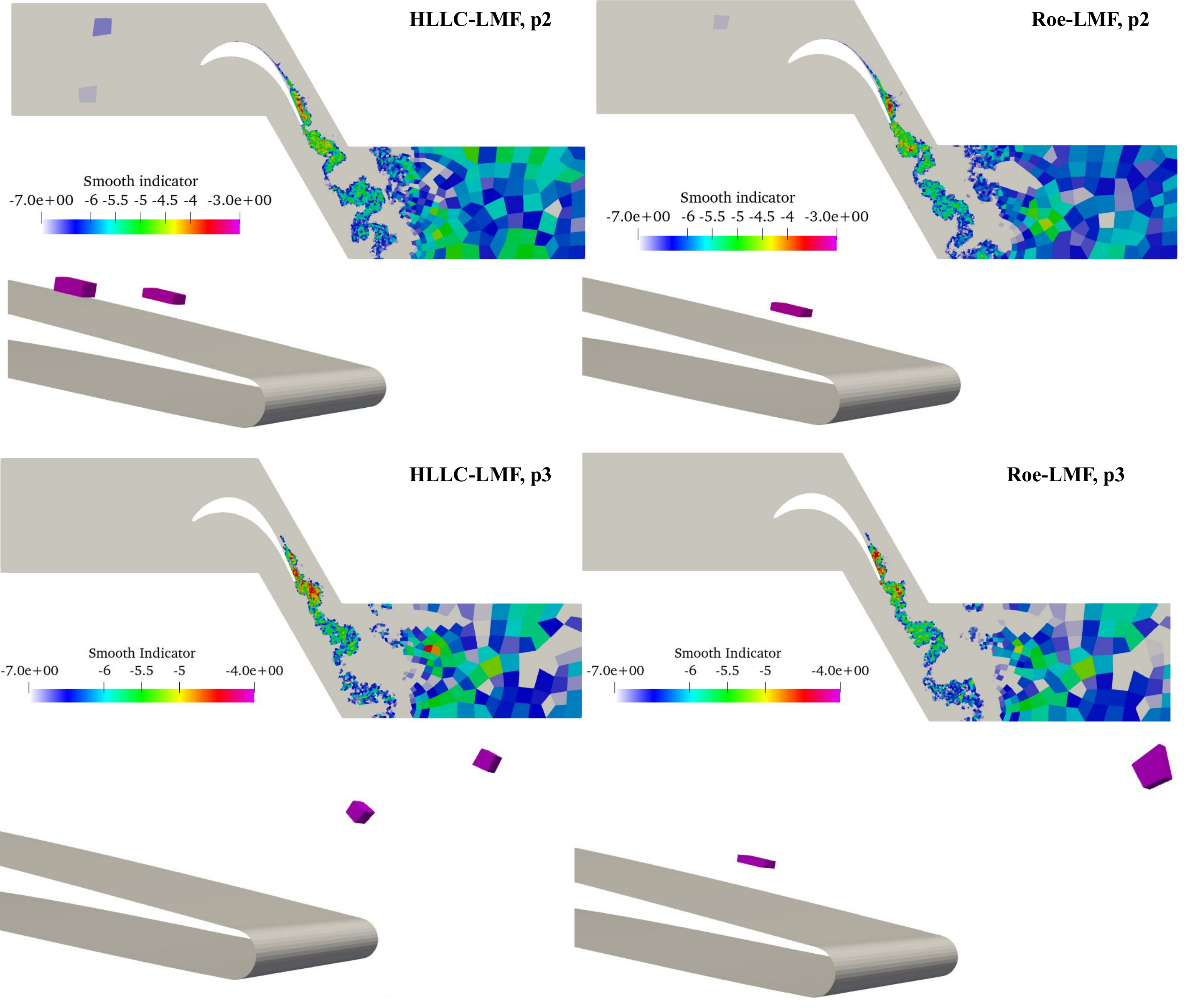}
\caption{Smooth indicator $s_e$ of two setups of the T106c turbine.}
\label{fig:t106c_sensor_new}       
\end{figure}

Figure~\ref{fig:t106c_sensor_new} shows the smoothness indicator $s_e$ of the instantaneous flow for HLLC-LMF and Roe-LMF for $p2$ and $p3$ computations. The $p2$ and $p3$ calculations use a value of -3 and -4, respectively. The flow regions with large flow gradients are successfully marked by $s_e$. For the p2 calculation, the transitional flow and wake regions are marked by values of approximately [-5:-3]. For $p3$, these regions are marked by a range of [-5.5:4]. For the flow upstream the leading edge, the flow is smooth and the value of $s_e$ is less than -7 for both $p2$ and $p3$. Therefore, choosing a value of less than -7 will essentially apply the modal filter to all cells. From a practical point of view, the order of accuracy is gradually raised to remove initial transient flows (e.g., from $p0$ to $p3$). A single value of $s_0$ could be used for this process, since the value of $s_0$ for a higher spatial order accuracy is normally a conservative value for a lower order computation. 

Figure~\ref{fig:t106c_sensor_new} also highlights cells that are subjected to filtering for instantaneous flow. It can be seen that only a handful of cells are actually filtered for both p2 and p3 computations. This confirms that LMF only targets certain local cells to stabilize the computation.


\subsubsection{The LS89 Turbine}
\begin{figure}
\centering 
\includegraphics[height=7cm]{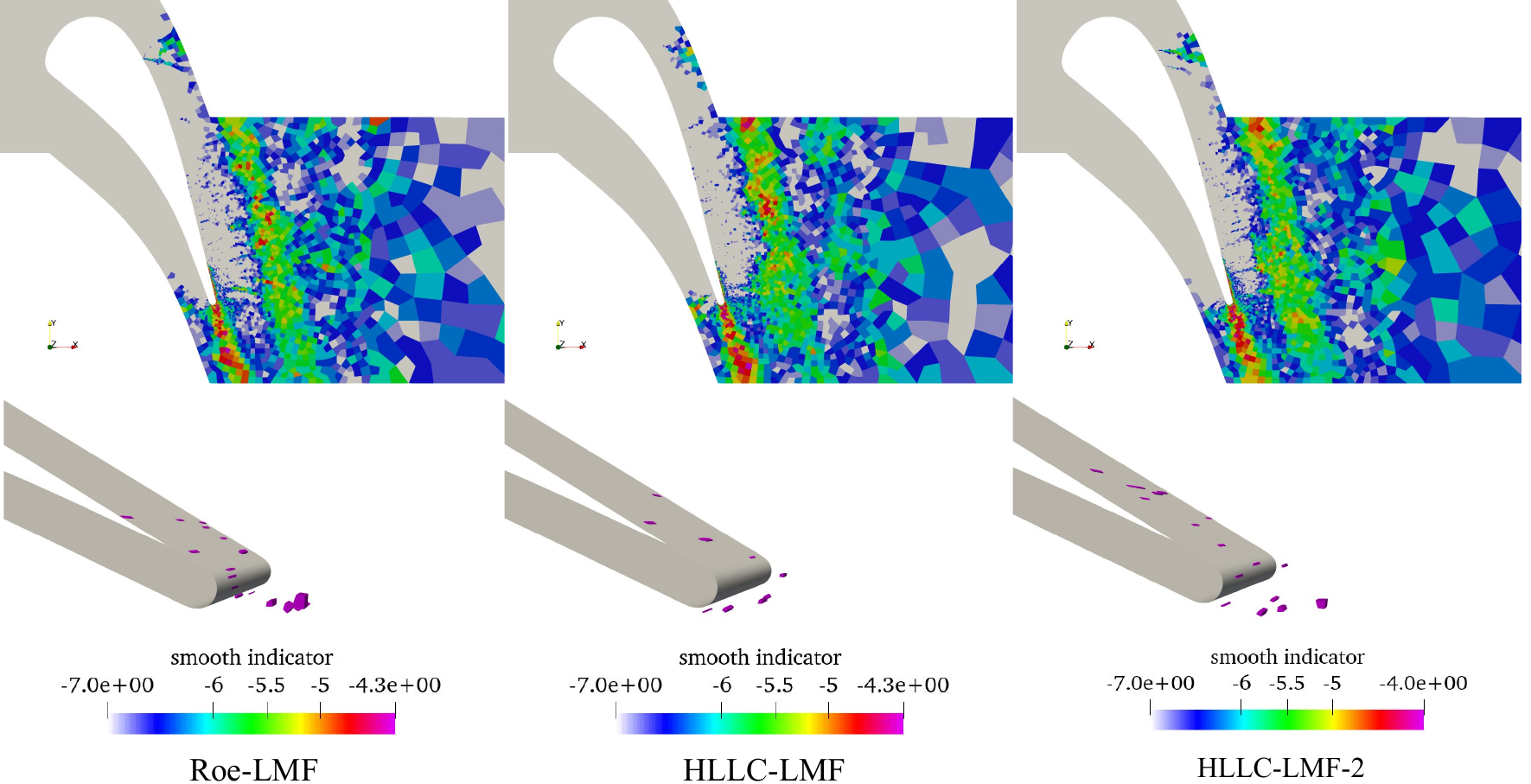}
\caption{Smooth indicator $s_e$ for different $p3$ setups of the LS89 turbine.}
\label{fig:ls89_sensor}       
\end{figure}

Figure~\ref{fig:ls89_sensor} shows the contours of the smooth indicator $s_e$ for the setups in Table~\ref{tab::ls89_setup} that use LMF. The plots in the bottom row highlight the cells that are subject to filtering.  From the plot, it can be seen that the smooth indicator can identify the regions where the flow is smooth and also the regions where the flows have a large spatial gradient. Compared to the T106c case, more cells are filtered due to the more complicated flow features in this case. However, only a tiny proportion of elements are filtered. This confirms that LMF introduces a very low dissipation. This demonstrates the effectiveness of the LMF in stabilizing the solution while leaving the smooth-flow region untouched. Furthermore, Fig.~\ref{fig:ls89_sensor} shows that the value of $s_e$ in the wake region and also in the transitional flow region is approximately in the range of $[-5:-4]$. This is consistent with the range suggested in Table~\ref{tab::smooth_indicator_value_new}.   

\begin{figure}
\centering 
\includegraphics[height=8.5cm]{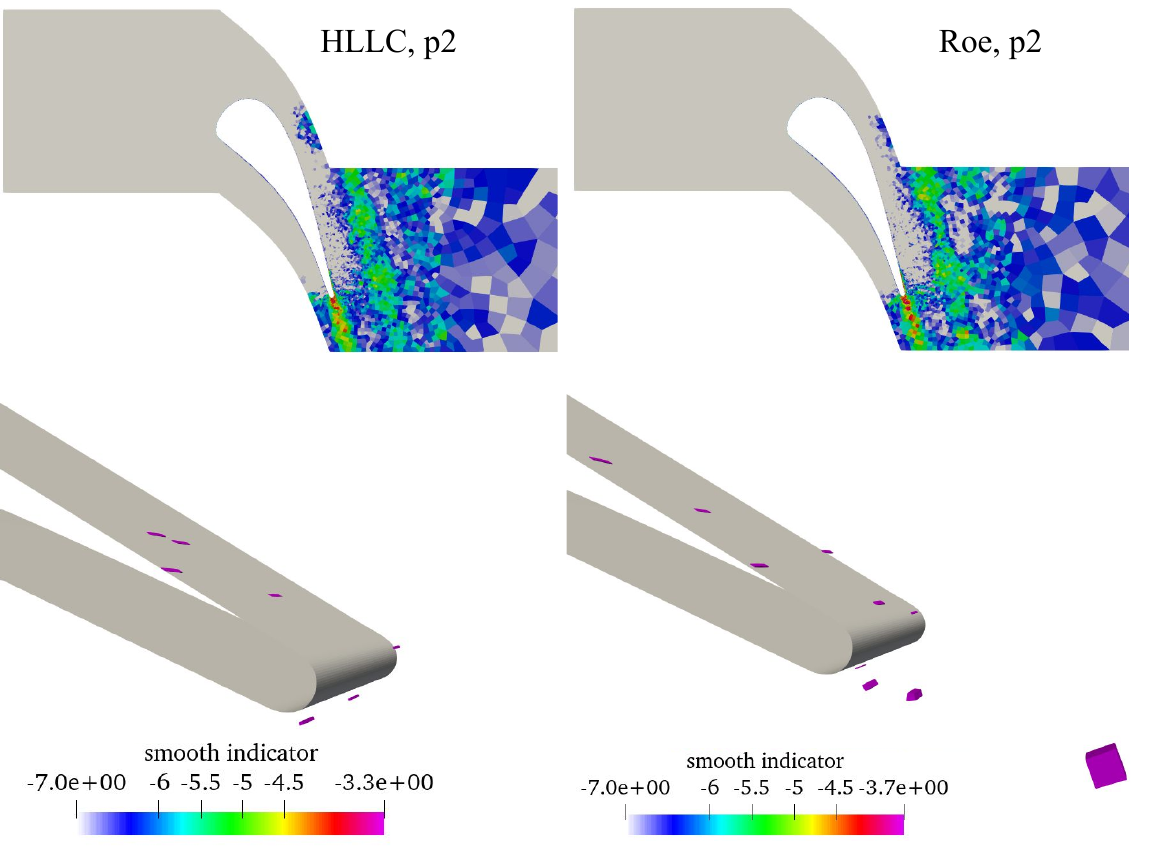}
\caption{Smooth indicator $s_e$ for different $p2$ setups of the LS89 turbine.}
\label{fig:ls89_p2_sensor}       
\end{figure}

The viability of the values in Table~\ref{tab::smooth_indicator_value_new} for a p2 computation of LS89 is demonstrated in Fig.~\ref{fig:ls89_p2_sensor}. The setups for the $p2$ computations are summarized in Table~\ref{tab::ls89_setup_p2}. Two sets of values of $s_0$ are attempted. Roe is found to be unable to stabilize with $s_0 = -3.3$ while HLLC can. This is consistent with the observations in the $p3$ computations. Therefore, in Fig.~\ref{fig:ls89_p2_sensor} HLLC has $s_0 =-3.3$ while Roe has the value of $-3.7$. 

\begin{table}
\centering
\caption{Different setups for the LS89 turbine $p2$ simulation}
\label{tab::ls89_setup_p2}
\begin{tabular}{cccccc}
\hline \hline
 Setup & order & $s_0$ & Riemann Solver & Filter & Success \\ \hline
 HLLC & p2 & -3.3 & HLLC &  LMF & \ding{51}  \\ 
 Roe & p2 & -3.3 & Roe & LMF & \ding{53}    \\
 HLLC & p2 & -3.7 & HLLC & LMF  & \ding{51}    \\
 Roe & p2 & -3.7 & Roe & LMF  &  \ding{51}  \\ \hline \hline
 \end{tabular}
\end{table}

From Fig.~\ref{fig:ls89_p2_sensor}, it can also be seen that the value of $s_e$ in the wake region and the transitional flow region is approximately in the range of $[-4.5:-3.3]$. This falls into the suggested values of $s_0$ for a p2 computation. 

\begin{figure}
\centering 
\includegraphics[height=4.5cm]{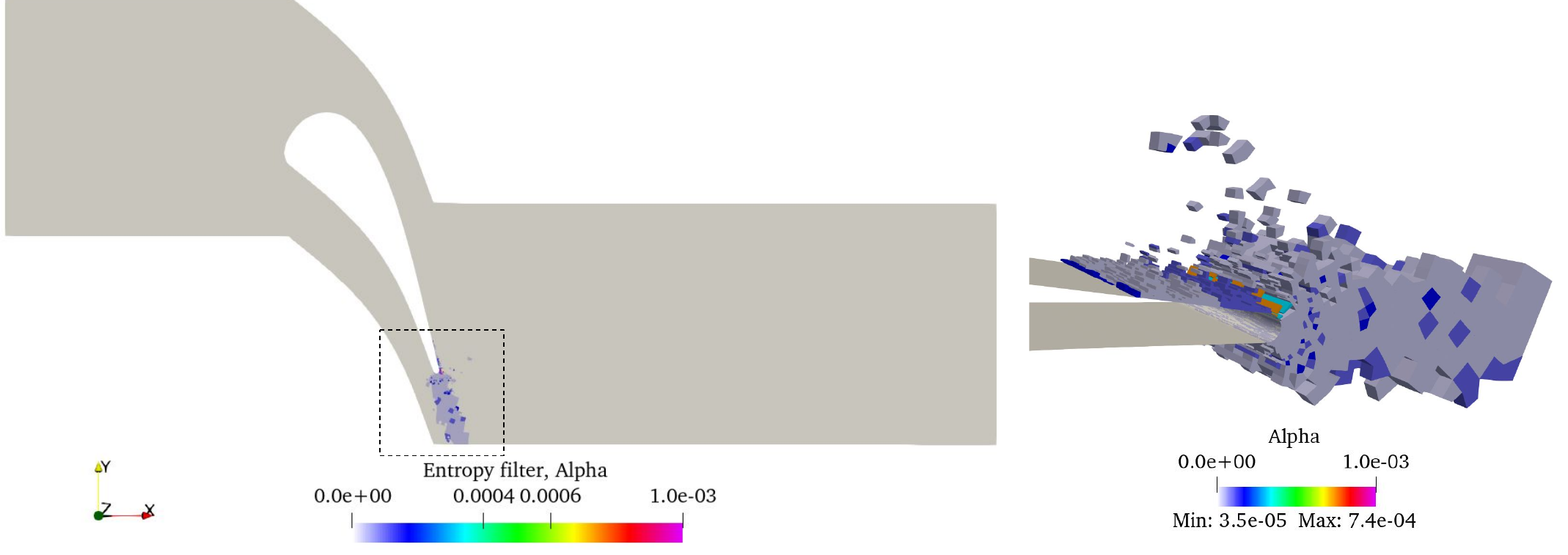}
\caption{Adaptively computed $\alpha$ in the second-order modal filter used in EF for the LS89 turbine .}
\label{fig:ls89_ef_alpha}       
\end{figure}

Regarding EF, $\alpha$ in the second-order modal filter used in EF is shown in Figure~\ref{fig:ls89_ef_alpha}. In EF, $\alpha$ is adaptively computed based on physical constraints (e.g., the minimum entropy principle and the positivity-preserving condition). When $\alpha=0$, no filtering is applied, it can be seen from Fig.~\ref{fig:ls89_ef_alpha} that the physical constraint in EF is capable of detecting the smooth flow region. When $\alpha > 0$, the second modal filter is applied to the element. Compared with Fig.~\ref{fig:ls89_sensor} and Fig.~\ref{fig:ls89_ef_alpha}, it can be seen that both EF and LMF target similar flow regions, but EF filters more elements than LMF, although the value of $\alpha$ is small for most elements that are subjected to filtering.  

\subsubsection{Remarks on LMF and EF}
As a comparison between LMF and EF, in terms of computational cost, EF requires an optimization process to obtain a suitable $\alpha$ in Equation~\ref{eqn::modal_filter_ef} for each cell. For LMF, only the smoothness indicator (see Equation~\ref{eqn::lmf_smooth_indicator}) needs to be computed, which is very efficient to evaluate. Therefore, LMF is computationally more efficient than EF. For the LS89 turbine case, each EF iteration is found to be roughly $15\%$ more expensive than that of LMF. In addition, since LMF is also easier to implement than EF, LMF can be a simple yet effective alternative to EF. On the other hand, EF is not completely parameter-free. It still has a tuning parameter ${\epsilon}_s$ (see Equation~\ref{eqn::ef_s}) and this parameter controls how strict the minimum entropy principle is preserved. More research is still required to understand how this parameter can be tuned to strike a balance between accuracy and stability. However, it should be noted that LMF and EF still follow the same strategy, which is to localize the modal filtering process to strike a balance between accuracy and robustness.

\section{Conclusions and Future Work}
An FR solver ATHOS has been developed and validated to simulate turbomachinery flows. Its performance is demonstrated in two industrially representative turbomachinery cases, and the results show excellent agreement with the experimental data. LMF has been developed to stabilize the solution, and the TGV case shows that LMF does not introduce a noticeable amount of numerical dissipation when the flow is smooth. A more relaxed spanwise grid spacing can be used to predict the HTC of the turbine blades. As the industry is more interested in mean flow quantities, this finding could lead to a reduction in the computational cost of industrial ILES simulations.  

The choice of Riemann solvers (i.e. HLLC or Roe) can have an impact on the non-linear stability of the solution; HLLC is found to be more robust than Roe. However, if stable results can be obtained, with respect to the mean flow only marginal differences are observed between the results of HLLC and Roe for the mean flow, and the predicted mean flow solution is dominated more by the smoothness criterion $s_0$ than by the choice of the Riemann solver; with respect to the instataneous flow, the PSD of pressure of the T106c turbine shows that the results from HLLC and Roe agree with each other in the inertial subrange but the PSD from the Roe solver produces a faster decay of PSD at higher frequencies. For the LS89 turbine case, all p3 solutions agree well with each other in the inertial subrange, a more dissipative stabilization approach, such as EF or the CPR limiter in Jia et al.~\cite{Jia2019} predicts a faster decay of PSD.

EF is able to stabilize the solution without tuning the parameters, but it can be more dissipative than LMF. Compared to EF, LMF is easier to implement and computationally cheaper than EF. Regarding the choice of the smoothness criterion $s_0$, the analysis shows that $s_0$ could be taken approximately from the range of $[-(p+2): -(p+1)]$, where $p$ is the order of polynomials (that is, 2 or 3). Therefore, LMF can be a simple but effective alternative to EF to stabilize FR simulations.

Future work includes exploring the possibility of using other flow variables to calculate the smoothness indicator. In order to further enhance the robustness, a more advanced approach could be used to compute the convective flux (such as an entropy stable scheme). Besides, the connection of EF or LMF to SVV within the FR framework could be explored, this could gain more insight into the behavior of EF and LMF and leads to potential future improvement. As was pointed out by Edoh et al.~\cite{Edoh2018}, solution-filtering approach can potentially lead to time inconsistency; this will also be investigated in future work.

\section*{Acknowledgments}
Calculations were performed using the Sulis Tier 2 HPC platform hosted by the Scientific Computing Research Technology Platform at the University of Warwick. Sulis is funded by EPSRC Grant EP/T022108/1 and the HPC Midlands+ consortium. The authors would like to acknowledge the open source projects FlurryPP~\footnote{https://github.com/JacobCrabill/FlurryPP}, HiFiLES\footnote{https://github.com/HiFiLES/HiFiLES-solver}, COOLFLUID\footnote{https://github.com/andrealani/COOLFluiD} and PyFR\footnote{https://github.com/PyFR/PyFR} for inspiring the current implementation of the high order flux reconstruction solver. The authors also would like to thank the reviewers for their helpful comments and suggestions in the revision of this manuscript. 

\appendix

\section{Overall Process of the FR implementation in ATHOS}
\label{sec:appendix_1}

The general process of the FR implementation in ATHOS is described in Algotirhm~\ref{alg::general_filtering}. Compared to the original FR method of Huynh~\cite{Huynh2007}, the addition is LMF and EF to stabilize the computation at each Runge-Kutta stage.
\begin{algorithm}
\caption{FR implementation in ATHOS}
\label{alg::general_filtering}
\begin{algorithmic}[1]
\Repeat
    \For{$i=1$ to $N_{RK}$}  \Comment{number of RK steps}
    \If{EF} 
      \State work out minimum entropy for each cell
    \EndIf
    
    \State FR algorithms \Comment{compute residuals and advance solution in time}
    
    \If{filtering} \Comment{LMF or EF}
      \If{EF} 
        \State perform the entropy filter for all the elements
      \ElsIf{LMF}
        \State Compute the smooth indicator $s_e$
        \ForAll{the elements}
           \If{$s_e \geq s_0$}
             \State Apply modal filtering for this element
           \EndIf
        \EndFor 
      \EndIf
    \EndIf
  \EndFor
\Until{iteration limit or wall-clock time limit}

\end{algorithmic}
\end{algorithm}

 \bibliographystyle{elsarticle-num} 
 \bibliography{elsarticle-template-num}





\end{document}